\documentclass{aa}  

\usepackage{graphicx}
\usepackage{lscape}
\usepackage{txfonts}
\usepackage[dvipsnames]{xcolor}
\usepackage{natbib}
\usepackage{hyperref}
\bibpunct{(}{)}{;}{a}{}{,}

\begin{document} 

\title{Protoplanetary and debris disks in the $\eta$\,Chamaeleontis Association:}
\subtitle{A sub-millimeter survey obtained with APEX/LABOCA\thanks{This publication is based on data acquired with the Atacama Pathfinder Experiment (APEX). APEX is a collaboration between the Max-Planck-Institut f\"ur Radioastronomie, the European Southern Observatory, and the Onsala Space Observatory. Based on 086.C-0174A-2010, PI: A. Sicilia-Aguilar; 082.F-9304-2008, PI: R.Liseau.}
 observations}

   \author{V.~Roccatagliata\inst{1,2,3}, A. Sicilia-Aguilar\inst{4}, M. Kim\inst{5,6},  J. Campbell-White\inst{7,4}, M. Fang\inst{8,9}, S. J. Murphy\inst{10}, S. Wolf\inst{11}, W. A. Lawson\inst{10}, Th. Henning\inst{12} \and  J. Bouwman\inst{12}}

\institute{\inst{1} INAF-Osservatorio Astrofisico di Arcetri, Largo E. Fermi 5, 50125 Firenze, Italy\\
\email{veronica.roccatagliata@inaf.it}\\
\inst{2} Department of Physics ``E. Fermi'', University of Pisa, Largo Bruno Pontecorvo 3, 56127 Pisa, Italy\\
\inst{3} INFN, Sezione di Pisa, Largo Bruno Pontecorvo 3, 56127 Pisa, Italy\\
\inst{4} SUPA, School of Science and Engineering, University of Dundee, Nethergate, DD1 4HN, Dundee, UK\\
\inst{5} Department of Physics, University of Warwick, Gibbet Hill Road, Coventry CV4 7AL, UK\\
\inst{6} Centre for Exoplanets and Habitability, University of Warwick, Gibbet Hill Road, Coventry CV4 7AL, UK\\
\inst{7} European Southern Observatory, Karl-Schwarzschild-Strasse 2, 85748 Garching bei M\"unchen, Germany\\
\inst{8} Purple Mountain Observatory, Chinese Academy of Sciences, 10 Yuanhua Road, Nanjing 210023, China\\
\inst{10} School of Science, University of New South Wales, Canberra, ACT 2600, Australia\\
\inst{11} Institute of Theoretical Physics and Astrophysics, University of Kiel, Leibnizstra\ss e 15, 24118 Kiel, Germany\\
\inst{12} Max Planck Institute for Astronomy, K{\"o}nigstuhl 17, D-69117 Heidelberg, Germany
}

\date{Received April 14, 2023; Accepted XX, 2023}

\abstract
{Nearby associations are ideal regions to study coeval samples of protoplanetary and debris disks down to late M-type stars. Those aged 5-10,Myrs, where most of the disk should have already dissipated forming planets, are of particular interest.}
{We present the first complete study of both protoplanetary and debris disks in a young region, using the $\eta$\,Chamaeleontis ($\eta$\,Cha) association as a test bench to study the cold disk content. We obtained sub-millimeter data for the entire core population down to late M-type stars, plus a few halo members.}
{We performed a continuum sub-millimeter survey with APEX/LABOCA of all the core populations of $\eta$\,Cha association. These data are combined with archival, multi-wavelength photometry to compile a complete spectral energy distribution. Disk properties have been derived by modeling protoplanetary and debris disks using RADMC 2D and DMS, respectively. We compute a lower limit of the disk millimeter fraction, which is then compared to the corresponding disk fraction in the infrared for $\eta$\,Cha. We also revisit and refine the age estimate for the region, using the Gaia eDR3 astrometry and photometry for the core sources.}
{We find that protoplanetary disks in $\eta$\,Cha typically have holes with radii of the order of 0.01 to 0.03\,AU, while ring-like emission from the debris disks is located between 20 au and 650 au from the central star. The parallaxes and Gaia eDR3 photometry, in combination with the PARSEC and COLIBRI  isochrones, enable us to confirm an age of $\eta$\,Cha between 7 and 9 Myrs. In general, the disk mass seems insufficient to support accretion over a long time, even for the lowest mass accretors, a clear difference compared with other regions and also a sign that the mass budget is further underestimated. We do not find a correlation between the stellar masses, accretion rates, and disk masses, although this could be due to sample issues (the objects are few and mostly low-mass). We confirm that the presence of inner holes is not enough to stop accretion unless accompanied by dramatic changes to the total disk mass content. Comparing $\eta$ Cha with other regions at different ages, we find that the physical processes responsible for debris disks (e.g., dust growth, dust trapping) efficiently act in less than 5 Myrs.}
{}


\titlerunning{APEX/LABOCA observations of the $\eta$\,Chamaeleontis Association} 
\authorrunning{Roccatagliata V. et al.}

\maketitle

\section{Introduction}

Protoplanetary disks are the site of planet formation and the source of the building blocks of forming planets \citep[e.g.,][]{Drazkowskaetal2023}. The last phase of the evolution of protoplanetary disks, the so-called debris disk phase, is characterized by remnant, second-generation dust without gas \citep[e.g.,][]{Wyatt2005}.  Recently, an increasing number of debris disks have been found to host gas \citep[e.g., ][]{Hughesetal2018}, which can be a leftover from primordial or a second-generation gas \citep[e.g.,][]{Kospaletal2013, Kraletal2017}. In particular, in the debris disks, only CO isotopologues have been detected  \citep[][]{Smirnov-Pinchukov+2022} most probably due to the self-shielding of CO itself. 

The dissipation of dust \citep[e.g.,][]{Hernandezetal2007, Sicilia-Aguilaretal2013} and gas \citep{Fedeleetal2010} in protoplanetary disks takes place within a few million years \citep[e.g.][]{Manaraetal2023} after the collapse of the molecular cloud. The major mechanisms responsible for the dissipation of the disk material are mass accretion \citep[e.g., ][]{hartmannetal2005,Manaraetal2023}, disk photoevaporation by the radiation of the central star or by external radiation  \citep{Alexanderetal2006a,owen10,Picognaetal2019}, planet formation \citep[e.g.,][]{Drazkowskaetal2023}, binarity \citep{Bouwmanetal2006}, and dynamical interactions \citep[e.g., ][]{Cuelloetal2019}.
The fundamental implication of these studies is that, in about 5 Myr, almost all the material required to form planets is gone, so planets must form within the first million years of disk evolution \citep[e.g.,][]{Sicilia-Aguilaretal2006a,Manaraetal2018}. 
Inside-out disk dissipation, which was long proposed as the expected mechanism for disk dispersal \citep{strom89}, was also thoroughly confirmed via observations with the Spitzer Space Telescope \citep{Sicilia-Aguilaretal2006a,Hernandezetal2007,teixeira12,Sicilia-Aguilaretal2013}, typically using a set of diagnostic diagrams in the infrared color-color plane \citep[e.g.,][]{Koepferletal2013}. The impact of high-mass stars in the dissipation timescale of protoplanetary disks is still controversial  \citep{Richertetal2018, Richeretal2015, Guarcelloetal2016, Gaczkowskietal2013, Roccatagliataetal2011}. 

The study of disk lifetimes, however, was not systematically extended to far-infrared and longer wavelengths, as highlighted, for instance, by \citet{ErcolanoPascucci2017}. This lack of 
completeness was caused by both the sensitivity of the instruments and the fact that not all members are identified in nearby associations.

At millimeter wavelengths, \citet{WilliamsCieza2011} and \citet{Williams2011} found a rapid evolution of disks within a few mega-years, interpreted as an efficient formation of millimeter-size grains rather than a decline in the disk mass. Recent ALMA large surveys, aiming mainly at Class II sources, led to a statistically significant analysis of the dust masses in clusters at different ages \citep[e.g., ][]{Williamsetal2019}, confirming a decrease of dust mass over time, as well as a positive correlation between disk and stellar mass \citep[e.g.,][]{Ansdelletal2016, Ansdelletal2018} and which is even steeper in older regions \citep{Pascuccietal2016}. ALMA observations of Class III sources in Ophiuchus \citep{Ciezaetal2019} and Lupus \citep{lovell2020LupusClassIII} suggested a rapid dispersal of millimeter-sized dust from protoplanetary disks and planetesimal belt formation already by 2 Myr. Protoplanetary disks resolved by ALMA, e.g., in the Disk Substructures at High Angular Resolution Project (DSHARP) disks \citep[][]{Stammler+2019}, triggered a new discussion on different evolutionary paths followed by resolved structured and not-structured Herbig \citep[e.g.][]{Garufi+2017} vs Tauri disks  \citep[e.g.][]{vanderMarel+2018}. Debris disks have been suggested to derive from the efficient dust traps in the planetesimal belts of structured disks \citep[e.g.,][]{Cieza+2021, Jiang+2021,Michel2021} while, for other disks, the decrease in dust mass would be mainly caused by radial drift. 

The evolution of protoplanetary and debris disks can be also strongly influenced by flybys \citep[e.g.][]{Cuelloetal2019, Cuelloetal2020}. In particular, \citet{Bertinietal2023} found a high incidence of close encounters for debris disks. This might have fundamental implications in the evolution of both the perturber source and the perturbed debris system since an exchange of material can take place during a close encounter \citep[e.g., as predicted by ][]{Picognaetal2014} and change in the inclination in the orbit of inner planets.

Nearby associations are ideal regions to draw complete samples of pre-main sequence stars and faint protoplanetary disks around stars with spectral types extending to late M-type stars. At about 100 pc and with an age between 5 and 10 Myr \citep[e.g., ][]{Luhmanetal2004}, the $\eta$\,Chamaeleontis ($\eta$\,Cha) association is unique in terms of coeval stars with spectral types between A and M6, different disk morphologies, accretion properties, and evolutionary stages. Its members have been studied via near-infrared ground-based data, {\it Spitzer} mid-infrared surveys \citep{Megeathetal2005, Sicilia-Aguilaretal2009} and {\it Herschel} far-IR data \citep{Riviere-Marichalaretal2015}. Initially discovered via X-ray \citep{Mamajeketal1999}, a deficit of low-mass stars in the core population suggested either a very compact formation scenario \citep{Morauxetal2007} or the presence of a halo or dynamically ejected cluster population. This halo population has been identified via accretion and stellar youth tracers as well as kinematic properties \citep{Lawsonetal2002,Luhmanetal2004,Lyoetal2004,Songetal2004,Murphyetal2010}. The total number of $\eta$\,Cha cluster members has been recently revised to be at least 23 using the eROSITA all-sky survey and Gaia eDR3 \citep{Robradeetal2021}. 

In this paper, we present a new sub-millimeter survey obtained with APEX to trace the emission of the outer part of disks and thus complement previous studies at shorter wavelengths. We also perform the first detailed analysis of disk properties for both protoplanetary and debris disks. In Sect.~\ref{obs} we summarize the Atacama Pathfinder Experiment (APEX)/Large APEX BOlometer CAmera (LABOCA) observations, data reduction, and the results, as well as the revised age of the association using the Gaia EDR3 photometric data and parallaxes. In Sect.~\ref{an} we analyze the spectral energy distributions (SED) of the protoplanetary and debris disks. Sect.~\ref{discussion} discusses the observed dust masses in the context of the mass accretion rates available in the literature, as well as the fraction of different classes of disks compiled from star-forming regions at different ages. A summary of our results and conclusions is given in Sect.~\ref{conclusions}.

\section{Observations and data reduction}
\label{obs}
The $\eta$\,Cha association was observed with APEX/LABOCA at 870 $\mu$m and with an angular resolution of 18.6 arcsec \citep{Siringoetal2009}.  The observations were carried out in {\it On-Off mode} (PHOT) mainly between the end of October and December 2010 and completed in November 2013. The data consists of a series of scans (and a variable number of sub-scans) and Tables~\ref{log_1} and \ref{log_2} summarize the corresponding total 
integration time. The opacity was determined with sky dips performed during the night, varying between $\tau$ = 0.09 and 0.41. The flux calibration was estimated by observing several calibrators per night, including Carina, CW-LEO, N2071IR, B13134, and PKS1057-79.

The on-off observations were calibrated and reduced using the LABOCA pipeline within the BoA package{\footnote{\href{Boa_pipeline}{www.apex-telescope.org/bolometer/laboca/reduction/BoA-woo.html}}}, developed for APEX bolometer data. BoA allows us to derive the opacity using the sky dips and to reduce the calibrators during each night.
The reduction of the On-Off data is obtained using the command {\it doOO} (included in the version of the BoA package of July 2010). This command reads as input a list of scans of the science target and the appropriate opacity of each scan. We used the {\it weak} option which allows the optimization of the reduction for faint sources. The reduction proceeds to compute the flux at each nodding phase. The final flux is the cumulative flux weighted with the flux error of each nodding. 
We also used the {\it clip} option to specify a sigma clipping as a threshold to remove the single noddings that fall outside it. The {\it clip} value adopted during the data reduction is 3. The final fluxes are reported in Table~\ref{fluxes}. The associated error, \,$\sigma$, is the statistical uncertainty measured on the data, computed as the standard deviation of all measurements, divided by the square root of the number of measurements. For non-detected sources, we report the 3\,$\sigma$ upper limit. 

To this dataset, we add archival\footnote{Program number: 082.F-9304A-2008, PI: R.Liseau} LABOCA maps of RECX 16, RECX 17 and RECX 18\footnote{Additional observations available in the archive with corrupted scans were not recoverable.}. Details of these observations are listed in Table~\ref{log_map}. BoA was also used to reduce the maps, in particular, the program {\it optimizes} for faint point sources (i.e., {\it reduce-map-weaksource.boa}). After the first iteration the signal to noise (S/N) over the reduced map was computed. To optimize the S/N, we performed three iterations of the data reduction. None of the three sources were detected. In Table~\ref{fluxes}, we report the 3\,$\sigma$ upper limit, where \,$\sigma$, for this observation mode, represents the standard deviation over the field covered by the map, without considering the edge and the bad pixels remaining in the map after the iterations.  Table~\ref{fluxes} lists the LABOCA fluxes and the astrometric measurements of parallax and proper motions compiled from GAIA eDR3 \citep{2021A&A...649A...1G}.

\begin{figure}
\centering
\begin{tabular}{c}
\includegraphics[width=0.98\linewidth]{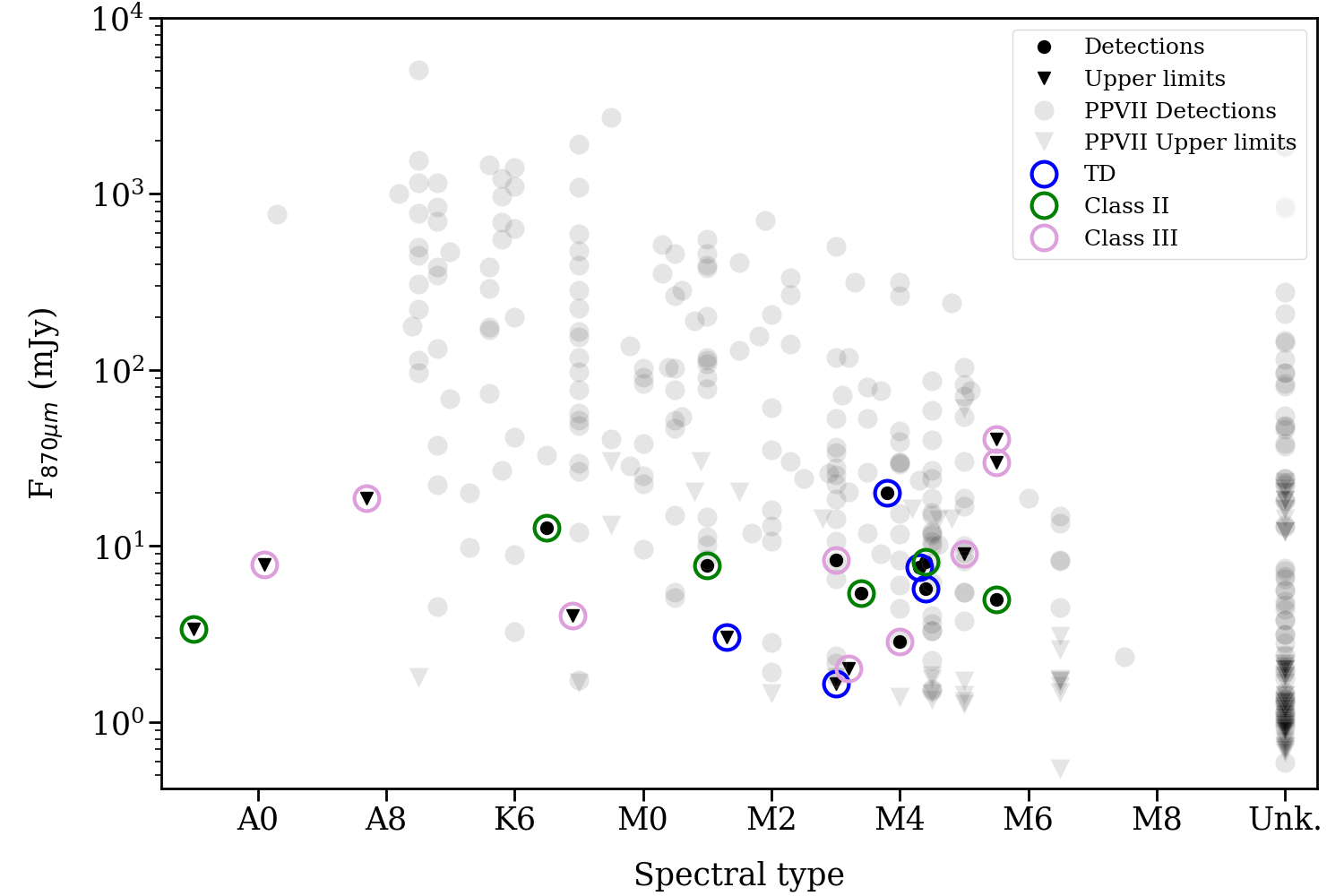} \\
\end{tabular}
\caption{APEX flux versus spectral type for the $\eta$\,Cha members. For comparison, the objects in the list from \citet{Manaraetal2023} are shown in grey, scaled to the distance of $\eta$\,Cha. Note the lack of significant correlations between these quantities. Filled dots represent flux detections, while inverted triangles show upper limits. A further color ring is added to specify the type of disk according to the classical SED classification adopted by \citet{Sicilia-Aguilaretal2009}: transitional disks (TD), Class II, and Class III (including the debris disks). RS Cha and $\eta$\,Cha are considered as upper limits due to cloud contamination.
\label{sptmdisk-fig}}
\end{figure}

\begin{table*}
\begin{small}
\caption{Summary of source properties and LABOCA measurements. Source names adopted in the paper and the corresponding SIMBAD database IDs are given in the first two columns, the spectral types in the third column, while parallaxes and proper motions from Gaia eDR3 are in columns four to six. The last column contains the LABOCA-measured fluxes and upper limits. The error in the fluxes is the root-mean-square (rms) of the observation. The upper limits are computed as $3\,\sigma$.} 
\label{fluxes}      
\begin{tabular}{r lrrr r r }  
\hline\hline                
\noalign{\smallskip}
Source      & SIMBAD&SpT & $\pi$&	$\mu_\alpha$	&$\mu_\delta$&	F$_{\rm 870 \mu m}$\\
&name&	&$[mas] $      & $[mas/yr] $ &$[mas/yr] $	&	[mJy]		\\
	\noalign{\smallskip}
\hline                
\noalign{\smallskip}	
{\bf Detections}&&&\\
\noalign{\smallskip}
\hline                
\noalign{\smallskip}
J0844	&	V* EO Cha                	&	M0             	&	10.1867	$\pm$	0.0106	&	-30.287	$\pm$	0.015	&	26.927	$\pm$	0.013	&	4.96	$\pm$	0.83	\\
J0843 	&	V* ET Cha                	&	M2             	&	9.6756	$\pm$	0.2249	&	-27.932	$\pm$	0.28	&	29.149	$\pm$	0.27	&	5.42	$\pm$	0.83	\\
RECX-5 	&	V* EK Cha                	&	M5             	&	10.1522	$\pm$	0.0138	&	-30.242	$\pm$	0.018	&	26.862	$\pm$	0.019	&	20.01	$\pm$	2.45	\\
RECX-6 	&	V* EL Cha                	&	M2             	&	10.1933	$\pm$	0.0156	&	-29.106	$\pm$	0.02	&	27.014	$\pm$	0.021	&	8.38	$\pm$	0.54	\\
RECX-8 $^1$	&	RS Cha                  	&	A7V            	&	10.137	$\pm$	0.0213	&	-27.263	$\pm$	0.032	&	28.179	$\pm$	0.029	&	18.62	$\pm$	2.63	\\
RECX-9 	&	V* EN Cha                	&	M4             	&	--		&		--		&		--		&	5.71	$\pm$	1.29	\\
RECX-11 	&	V* EP Cha                	&	K5             	&	10.1198	$\pm$	0.0111	&	-30.145	$\pm$	0.015	&	26.801	$\pm$	0.011	&	12.74	$\pm$	1.68	\\
\noalign{\smallskip}
J0801 $^2$ 	&	2MASS J08014860-8058052  	&	M4.4           	&	8.9432	$\pm$	0.0233	&	-19.006	$\pm$	0.031	&	28.879	$\pm$	0.03	&	7.58	$\pm$	0.86	\\
J0820 $^2$ 	&	2MASS J08202975-8003259  	&	M4.3           	&	10.2685	$\pm$	0.0229	&	-27.872	$\pm$	0.028	&	29.282	$\pm$	0.032	&	8.12	$\pm$	1.1	\\
RX\_J1005 $^2$ 	&	RX J1005.3-7749          	&	M1e            	&	10.0257	$\pm$	0.0129	&	-37.326	$\pm$	0.016	&	14.656	$\pm$	0.015	&	7.75	$\pm$	0.56	\\
\noalign{\smallskip}
\hline                
\noalign{\smallskip}	
{\bf Marginal}&{\bf Detections}&&\\
\noalign{\smallskip}
\hline                
\noalign{\smallskip}
$\eta$\,Cha	&	* eta Cha        	&	B8V            	&	10.1679	$\pm$	0.0665	&	-29.43	$\pm$	0.18	&	26.831	$\pm$	0.119	&	3.37	$\pm$	1.86	\\
J0841 	&	V* ES Cha                	&	M4             	&	10.1205	$\pm$	0.0247	&	-29.957	$\pm$	0.034	&	28.036	$\pm$	0.036	&	2.88	$\pm$	0.97	\\
\noalign{\smallskip}
\hline                        
\noalign{\smallskip}
{\bf Upper} &{\bf limits}&&\\
\noalign{\smallskip}
\hline                
\noalign{\smallskip}
RECX-3 	&	V* EH Cha                	&	M3             	&	10.1168	$\pm$	0.0107	&	-28.442	$\pm$	0.014	&	27.001	$\pm$	0.016	&	<	1.65		\\
RECX-4 	&	V* EI Cha                	&	K7             	&	10.1366	$\pm$	0.0121	&	-30.826	$\pm$	0.016	&	26.006	$\pm$	0.018	&	<	3.03		\\
RECX-7 	&	V* EM Cha                	&	K3             	&	10.1278	$\pm$	0.0118	&	-30.184	$\pm$	0.014	&	27.576	$\pm$	0.014	&	<	4.02		\\
HD 75505 	&	HD 75505                 	&	A1V      	&	10.2039	$\pm$	0.0213	&	-30.646	$\pm$	0.028	&	27.134	$\pm$	0.031	&	<	7.86		\\
RECX-12 	&	V* EQ Cha                	&	M3        	&	10.2837	$\pm$	0.0354	&	-30.397	$\pm$	0.045	&	26.256	$\pm$	0.041	&	<	2.01		\\
RECX-16 	&	2MASS J08440915-7833457  	&	M5.5       	&	10.0441	$\pm$	0.0393	&	-29.844	$\pm$	0.054	&	26.262	$\pm$	0.049	&	<	40.4		\\
RECX-17 	&	ECHA J0838.9-7916        	&	M5         	&	10.0827	$\pm$	0.0363	&	-28.868	$\pm$	0.044	&	27.035	$\pm$	0.041	&	<	9.10		\\
RECX-18 	&	ECHA J0836.2-7908        	&	M5.5       	&	9.9086	$\pm$	0.05	&	-29.092	$\pm$	0.059	&	27.871	$\pm$	0.053	&	<	29.8		\\
\noalign{\smallskip}
\hline                        
\noalign{\smallskip}
{\it Herschel}&{ }&&&&&\\
\noalign{\smallskip}
\hline                
\noalign{\smallskip} 
RECX1&V* EG Cha  &K4Ve           	& 10.1099	$\pm$	0.0917	&	-29.62	$\pm$	0.118	&	26.852	$\pm$	0.103	&\\ 
 \noalign{\smallskip}
\hline                        
\noalign{\smallskip}
\end{tabular}
\\$^1$: The detection of RECX-8 is contaminated by a local extended nebulosity highlighted by {\it Planck} between 30 an 217 GHz. For this reason, it is more appropriate to consider the emission from RECX-8 as an upper limit. 
\\$^2$: Halo members from the work of \citet{Murphyetal2010}.\\
\end{small}
\end{table*}

\section{Analysis}
\label{an}

In this section, we provide the details of the data analysis. After computing the dust disk mass from the sub-millimeter flux, the spectral energy distributions (SEDs) of the cluster members were compiled, including our new submillimeter photometric point from our APEX/LABOCA survey as well as archival optical magnitudes, 2MASS JHK$_s$ data \citep[][]{Sicilia-Aguilaretal2009}, near-infrared (WISE and Akari), mid-infrared observations from {\it Spitzer} \citep{Sicilia-Aguilaretal2009} and far-infrared data from {\it Herschel} \citep{Riviere-Marichalaretal2015}.

So far, only the dust composition in the inner part of the disk, observed by {\it Spitzer}/IRS, has been analyzed for the $\eta$\,Cha targets \citep{Bouwmanetal2006,Sicilia-Aguilaretal2009}, together with a first interpretation of the SEDs including up to mid-infrared wavelengths. The SEDs including the {\it Herschel}/PACS photometry  \citep{Riviere-Marichalaretal2015} were not modeled, which we now do including the submillimeter data. Our LABOCA survey did not cover RECX 1, but we add its {\it Herschel} data to our analysis.  

We then analyze the structure of protoplanetary disks and debris disks in the $\eta$\,Cha association fitting the complete SEDs. Protoplanetary disks are fitted using the Monte Carlo radiative transfer code RADMC 2D \citep{dullemond04,dullemond11}, while for debris disks we use the DMS \citep[Debris disks around Main sequence Stars;][]{Kimetal2018} software, optimized for optically thin emission.  For 2MASS J08014860, since its nature is not clear, we proceed with the modeling using both codes, RADMC 2D and DMS.
 
\subsection{Dust disk masses}
In Fig.~\ref{sptmdisk-fig} we show the fluxes as a function of the spectral type of the star, noting that we detect sources down to late M-type stars. For comparison, we also show the sources with known spectral types in the list from \citet{Manaraetal2023}, scaled to the distance of $\eta$\,Cha. We highlight that there is no correlation between the fluxes detected and the spectral type of the host star. A Spearman rank test returns a false-alarm probability of 12\% if we consider only the detections, or 65\% if upper limits are included, which is fully consistent with the two quantities being uncorrelated. We highlight that our work is homogeneously tracing different spectral types in various regimes only barely covered in other works. 

Under the assumption that the emission at millimeter wavelengths is optically thin, the dust disk mass ($M_{\rm dust}$) is directly proportional to the millimeter flux ($S_{\nu}$), so
   \begin{equation}\label{mdust-eq}
     M_{\rm dust}=\frac{S_{\nu}D^2}{k_{\nu}B_{\nu}(T_{\rm dust})},
   \end{equation}
where $ k_{\nu}=k_0(\nu/\nu_0)^{\,\beta} $ is the mass absorption coefficient, $\beta$ parametrizes the frequency dependence of $k_{\nu}$, $S_{\nu}$ is the observed flux,
$D$ is the distance to the source, ${\it T}_{\rm dust}$ is the dust temperature, and $B_{\nu}(T_{\rm dust})$ is the Planck function. For consistency, we adopt the dust properties from \citet{Ansdelletal2016}, using typical assumptions of a single dust grain opacity 
$k_0$ = 3.37 cm$^2$/g at 890\,$\mu$m and a single dust temperature $T_{\rm dust}$ = 20 K, which corresponds to the median temperature of the Taurus disks \citep{AndrewsWilliams2005}. This approach allows us to directly compare our results with those of \citet{manara16}. 
In this calculation, we have always highlighted that any trend in the dust mass can reflect the variation in opacity or the differences in temperature among the variety of disks in the cluster. Moreover, as already highlighted by \citet{Hartmannetal2006}, disk masses are systematically underestimated by a factor that could be up to hundreds \citep[e.g.,][]{Liu+2022,Rilinger_2023}, due to the uncertainties in the factual dust properties, disk size, and inclination. 
   
The fact that most disks from \citet{Manaraetal2023} are substantially more massive/more luminous than the $\eta$~Cha ones is a signature of a bias and suggests that the understanding of protoplanetary disks and their dissipation timescales and mechanisms requires including the faint, harder-to-detect, more evolved targets as well. This is not the only bias present, since until a few years ago, debris disks were more commonly detected around early-type stars \citep[e.g.,][]{Matthews+2014, Sibthorpe+2018} basically reflecting a skew in the observed sample. As \citet{vanderMarel+2021} started to highlight, the observed debris and structured disks show instead similar detection rates as a function of their host stellar mass. 

\subsection{Modelling the protoplanetary disks with RADMC-2D}

To study the structure of the protoplanetary disks in the $\eta$\,Cha association, we fitted the full SED using the Monte Carlo radiative transfer code RADMC 2D \citep{dullemond04,dullemond11} for different underlying disk structures to explore the various possibilities for each system. RADMC 2D assumes a fully irradiated disk where the temperatures of all the dust species for each given location are the same, which imposes the main limitation on the models. The models can be set to have well-mixed gas and dust (iterating on the solution and using the local temperature to establish the flaring), or decoupled gas and dust (setting a given power law for the vertical scale height $H$ vs. disk radius $R$) to simulate a more settled disk, even though all species would be settled in the same way, which is a simplification. The stellar photospheres rely on the same MARCS models \citep{gustafsson08}  used in \citet{Sicilia-Aguilaretal2009}, with a temperature and radius that reproduce the stellar SED for each object. Interstellar extinction is considered to be negligible, which is well supported by an independent analysis carried out by \citet{Rugeletal2018} of X-Shooter spectra of the $\eta$\,Cha core members. The stellar properties are summarized in Table \ref{fluxes}. 

We follow the same strategy as in \citet{Sicilia-Aguilaretal2015} where, rather than trying to derive the best model, we try to understand whether each disk can be reproduced by a standard, continuous model, or if it requires radial (e.g., variations in composition, inner holes or gaps) or vertical changes (e.g., variations in the disk scale height in the inner and outer disk, settling and deviations from a well-mixed dust and gas disk). 
We explore different disk hole sizes and vertical scale heights to reproduce the observed disk emission.
Taking into account that the parameter space is non-continuous and highly degenerated even for relatively complete SEDs with Herschel and millimeter data \citep{Sicilia-Aguilaretal2015,sicilia16}, and the variety of processes that can affect disks, the best fit is rarely unique (and often not perfect), but even in this case it is possible to assess whether a disk presents signs of evolution.
Since most of the SEDs show clear evidence of holes/gaps and a radial variation in the dust properties between the inner and the outer disk, we concentrate on exploring the different possibilities for inner vs. outer disk structure to reproduce the observed SEDs. This includes changes in composition, flaring, vertical scale height, and imposing a given H/R law that simulates dust setting vs. flared gas in hydrostatic equilibrium. Because of the complexity of the systems and the lack of resolved observations, it is often impossible to clearly distinguish between scenarios, but deviations with respect to standard and less evolved disks are evident.

\begin{figure*}
\centering
\begin{tabular}{cc}
\includegraphics[width=0.38\linewidth]{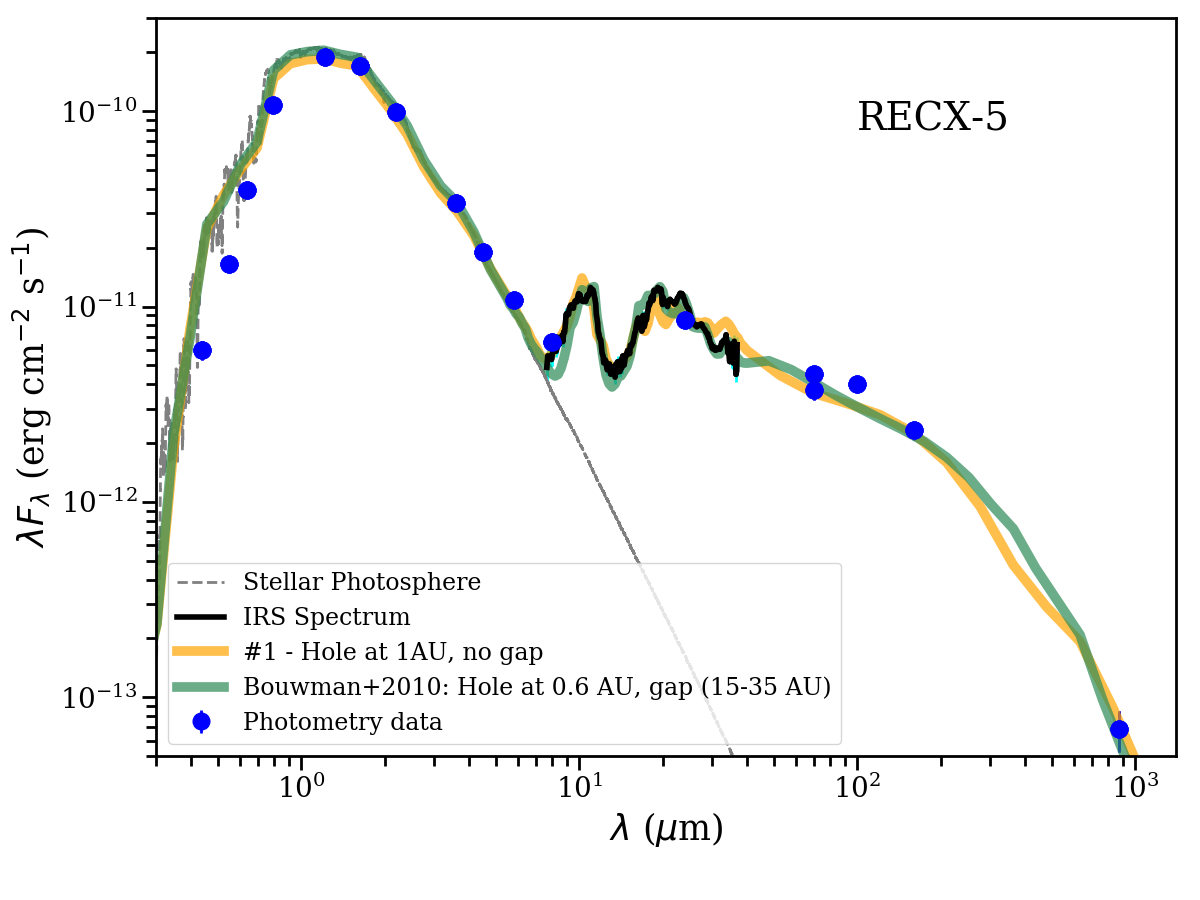} &
\includegraphics[width=0.38\linewidth]{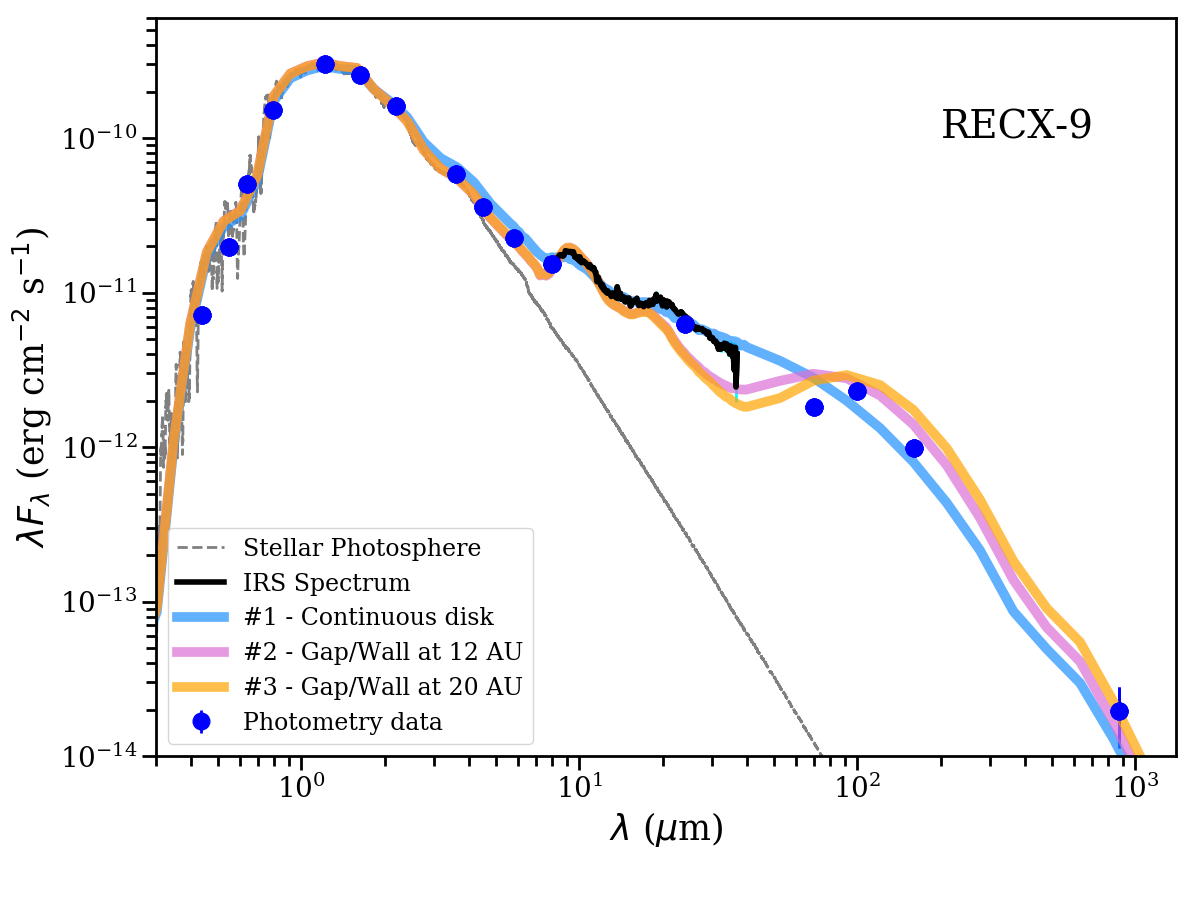} \\
\includegraphics[width=0.38\linewidth]{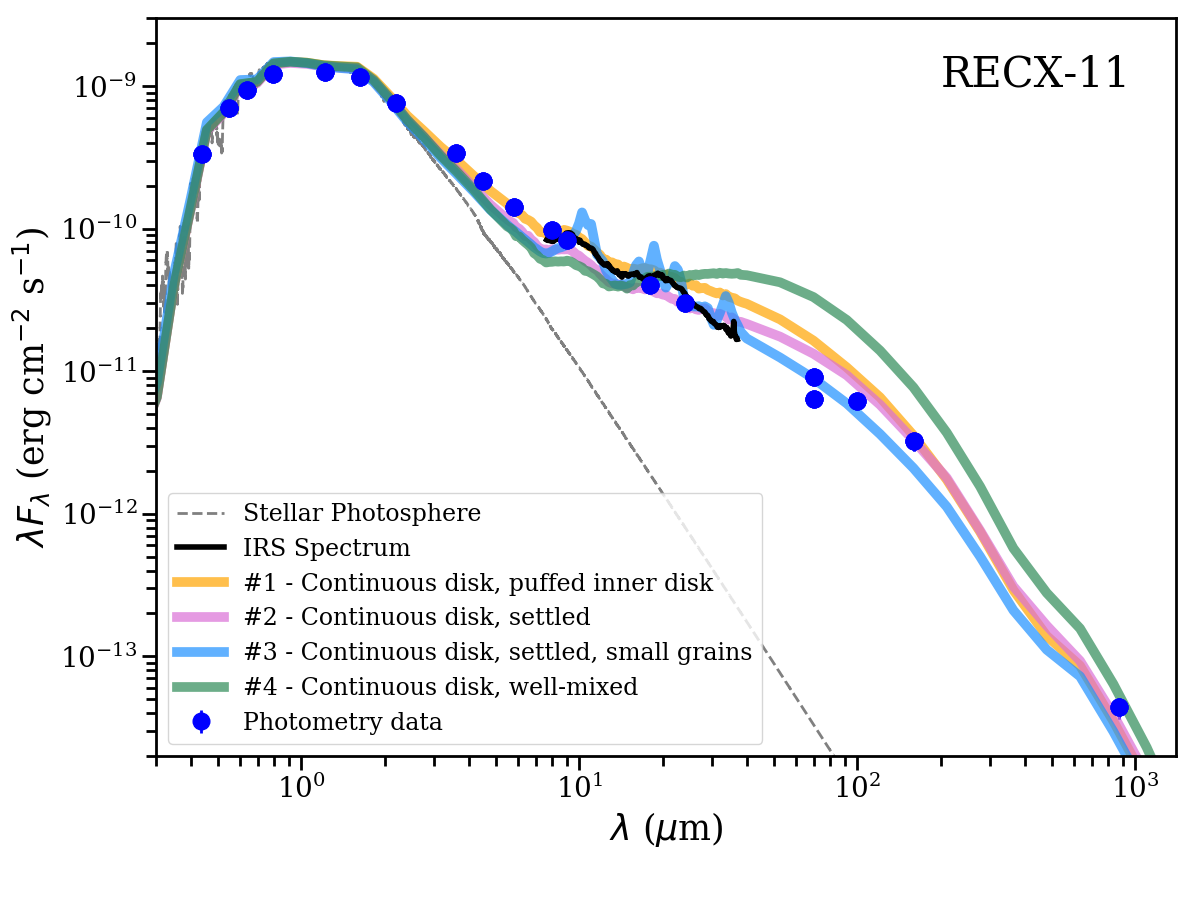} &
\includegraphics[width=0.38\linewidth]{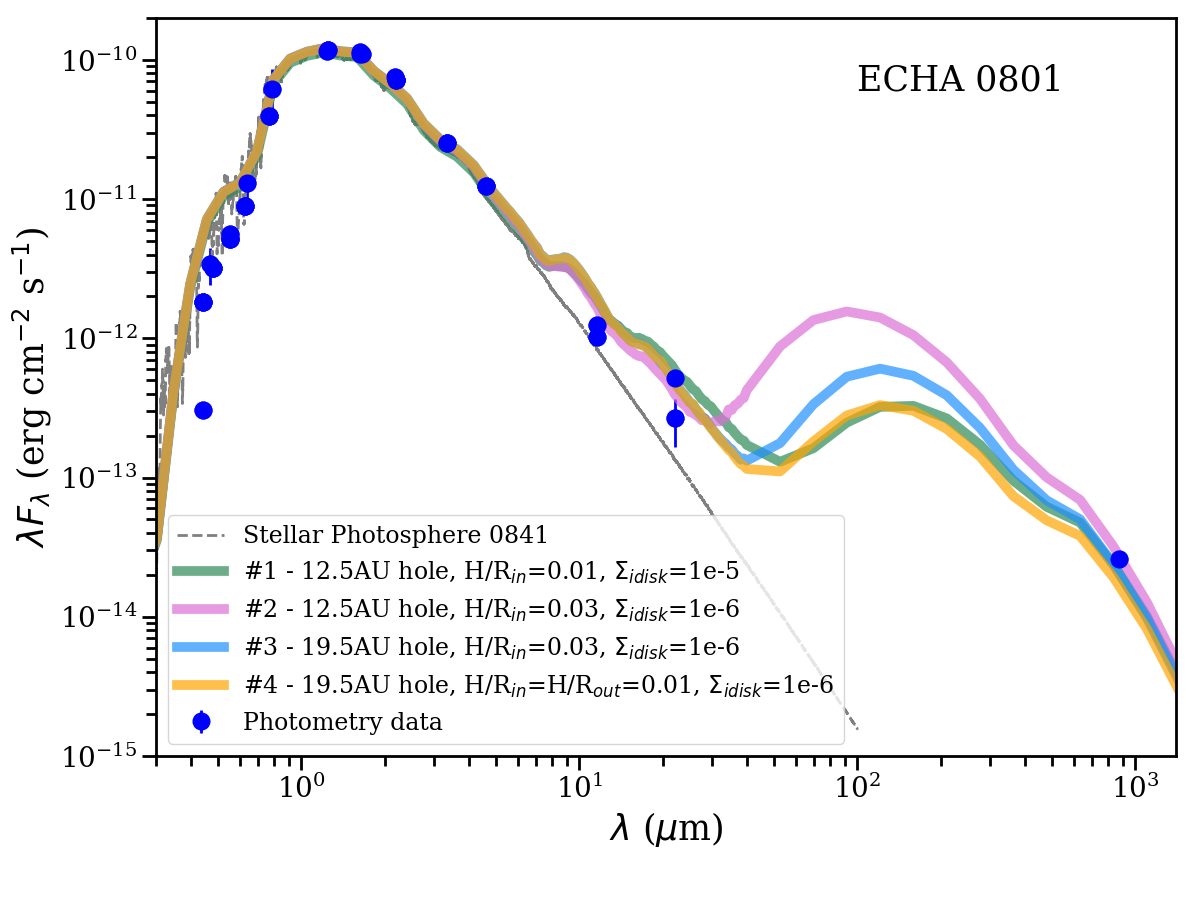} \\
\includegraphics[width=0.38\linewidth]{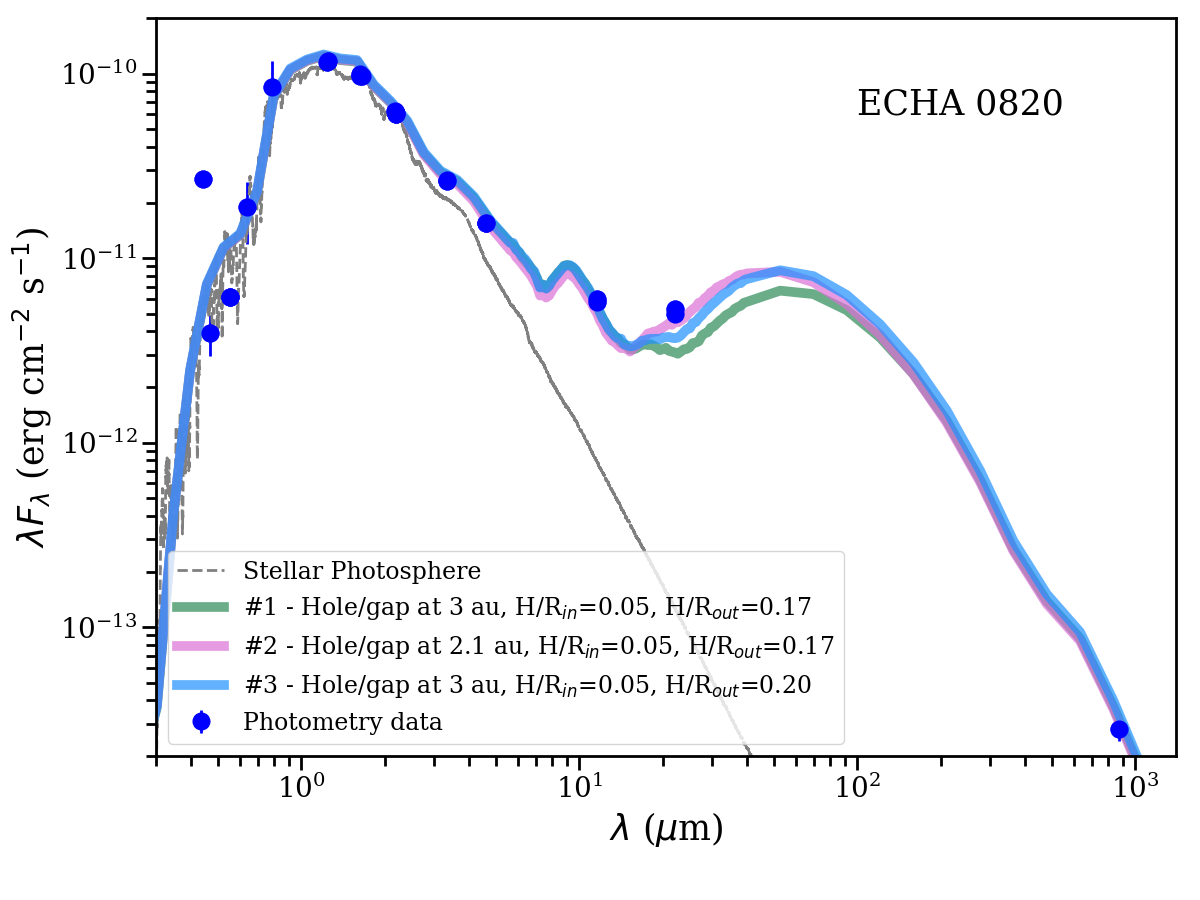} &
\includegraphics[width=0.38\linewidth]{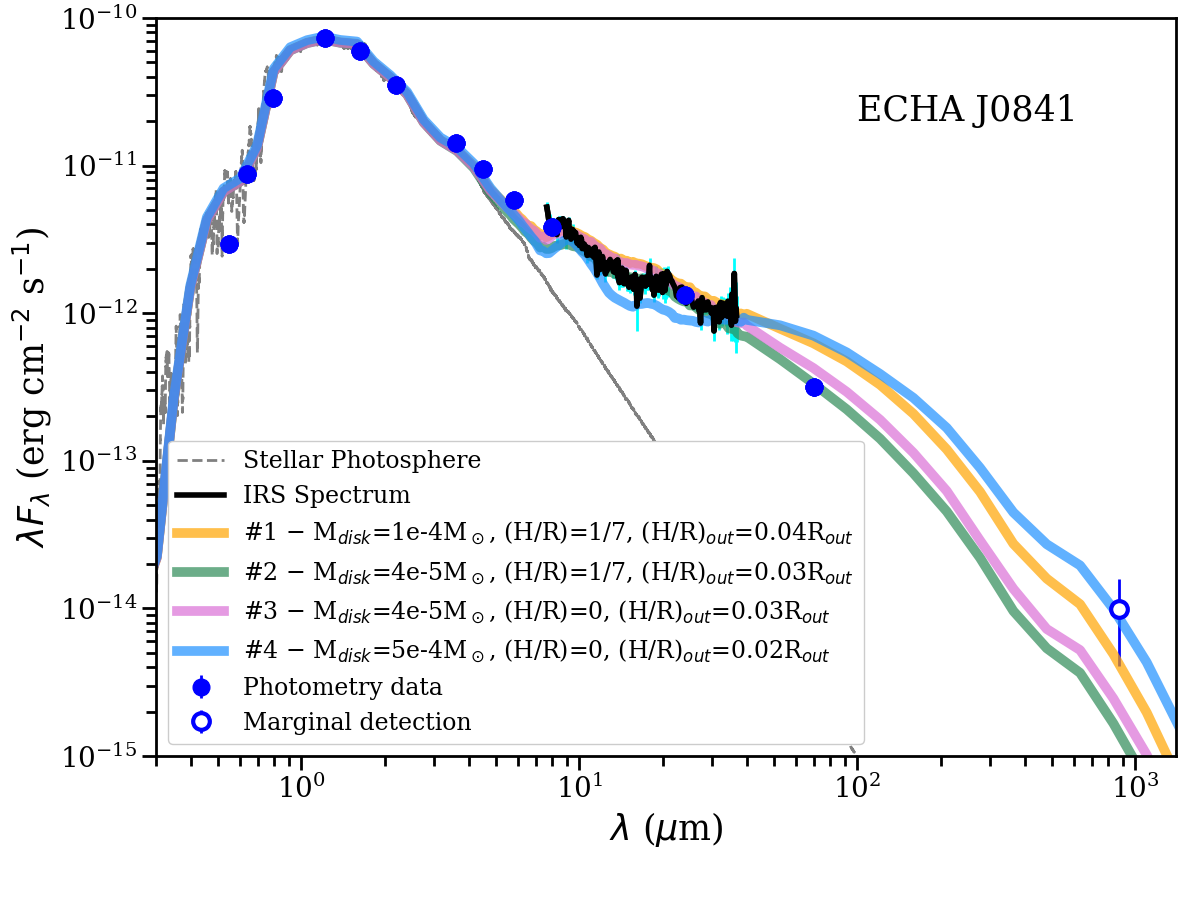} \\
\includegraphics[width=0.38\linewidth]{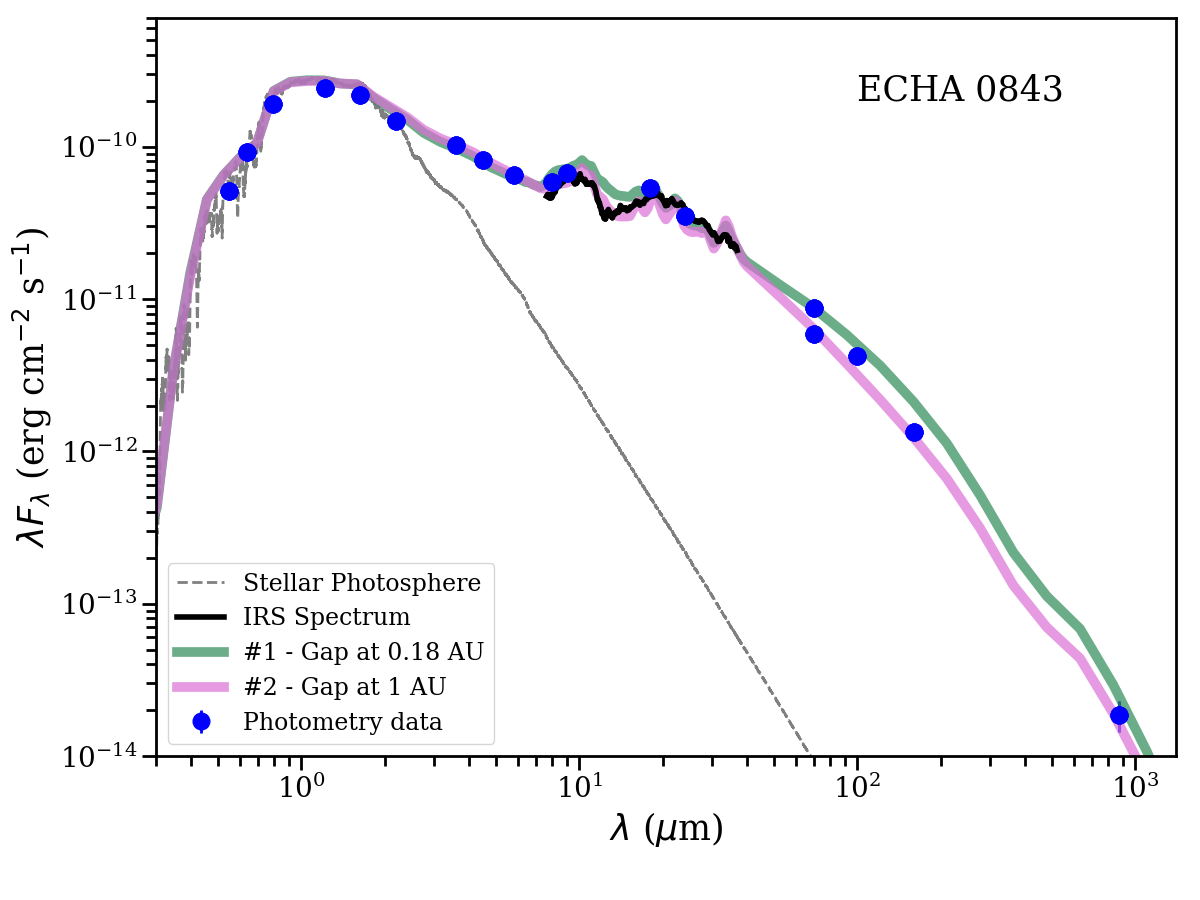} &
\includegraphics[width=0.38\linewidth]{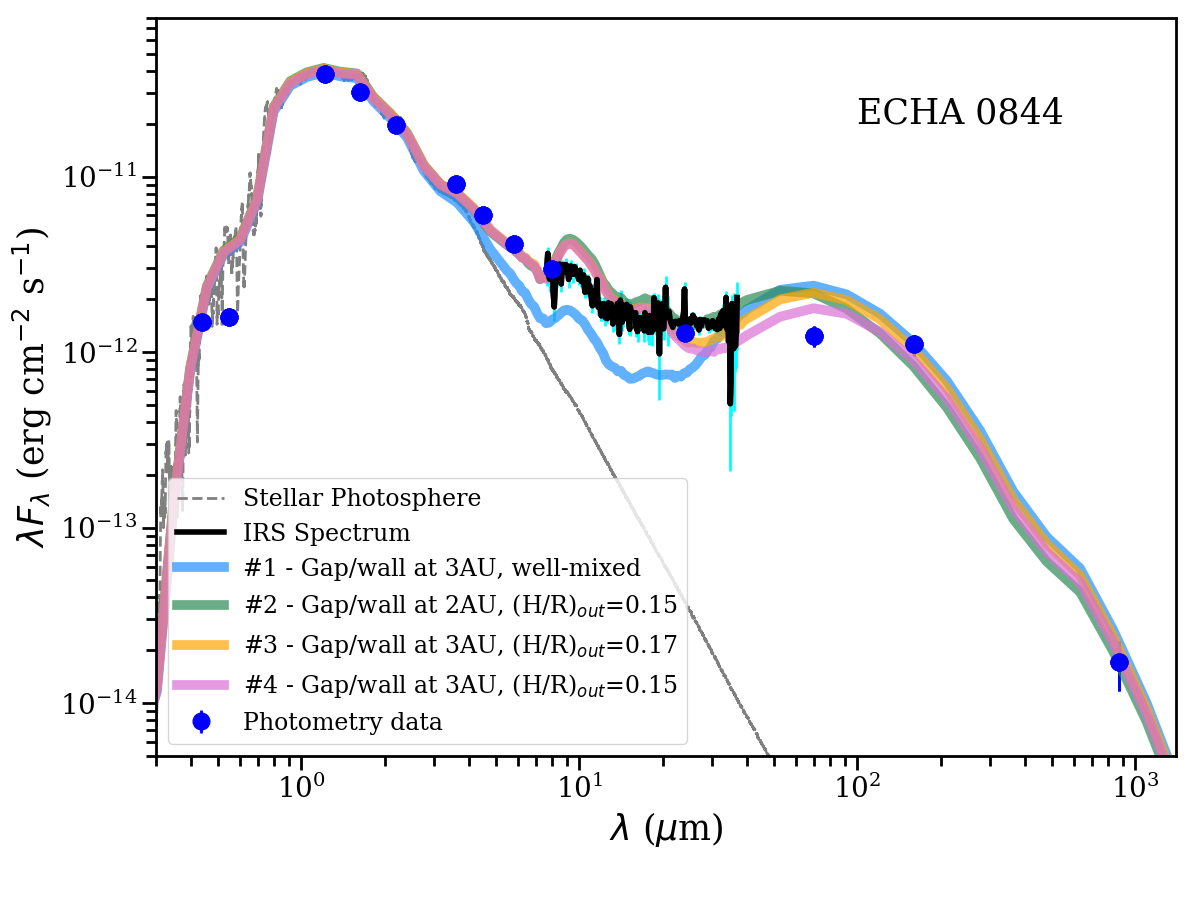} \\
\end{tabular}
\caption{SED models for the protoplanetary disks. For all cases, the photometry data is marked by blue circles (empty symbols for marginal detections). Spitzer/IRS spectra are plotted in black. The MARCS model stellar photospheres are marked by dotted grey lines, and the various disk models are represented by colored lines and labeled according to their main characteristics (see Table \ref{models-table}). RECX-5: A large-scale gap is not needed to reproduce the long wavelengths, which also means that SED alone cannot constrain among many diverse but equally plausible SED structures. RECX-9: A change in vertical scale height at 10-15 au is needed, which could be caused by a gap, wall, warp, or any other structure affecting the density and the scale height probably created by the existing companion at 20 au. RECX-11: Best fit with relatively massive and flattened disks. A more puffed innermost disk (either a puffed-up rim or a more extended $\sim$0.6 au region) is required, with the disk becoming increasingly flattened and settled at larger radii. J0801 and J0820 appear to be examples of relatively massive transition disks with large, strongly mass-depleted inner holes. ECHA J0841: Very flattened SED. ECHA J0843: Small gap or hole required. For ECHA J0844: Gap and/or change in the vertical scale height needed to explain the far-IR flux. \label{models-fig}}
\end{figure*}

For the dust model, we used a collisional distribution of amorphous magnesium and iron silicate grains with sizes between 0.1 and 1000 $\mu$m and a 25\% content of amorphous carbon. The dust opacities were taken from the Jena database \citep{jaeger03}. In the innermost disk (inside the hole, which does not necessarily need to be completely clean but may have lower densities), we can keep the same dust distribution as in the outer disk, or consider an optically thin population composed of 0.1 $\mu$m grains with a mixture of amorphous silicates and crystalline forsterite in a 1:3 proportion. Small and crystalline grains in the inner disk are included depending on whether the object has a strong silicate feature. 
Large grains in the inner disk tend to make the silicate feature fade, although technically it is very hard to distinguish a radial vs. vertical (e.g., due to differential settling) stratification of large and small grains. For disks with weak silicate emission, keeping the same composition for the inner and outer disk can typically reproduce the observed SED, so we chose to use a uniform grain distribution (with variable density and vertical scale height, if needed) for simplicity. The interface between the inner and outer disk composition is set using the Schmidt number (Sc), for which larger values imply a more abrupt composition change. In general, we take Sc=2/3 to ensure a smooth transition, but this may be increased to simulate more abrupt gaps. The gas-to-dust ratio is set to 100.

By default, the outer disk (or the whole disk, if no distinct radial substructures are included) is considered to be flared with $H/R$ $\propto$ $R^{\rm \,1/7}$, where $H$ is the vertical scale height of the disk and $R$ is the disk radius. The value of H/R in the outer part of the outer ($H/R_{\rm \,out}$) and inner ($H/R_{\rm \,in}$) disk, and the minimum vertical scale height can be changed to reproduce the observed SED.
Without resolved observations, there is a strong degeneracy between the disk mass and the outer disk radius, even in the presence of submillimeter data. We thus take the disk radius to be R$_{out}$ 100 au. The inclination angle is taken to be 45 degrees in all cases, noting that the most dramatic differences occur only at extreme inclinations. 
For the inner disk, when present, we set up an inner and outer region, and a surface density that changes with a certain power of the radius, typically taken to be -1. The inner disk flaring is set by default to $H/R$ $\propto$ $R^{\rm \,1/7}$, except in very flattened models. To avoid overloading the models with extra parameters, we do not include extra puffed inner rims, although if the height of the inner and outer disks do not match, this effectively behaves as a puffed rim.

Since the parameter space is degenerated, we investigated the minimal model needed to account for the observed SED, rather than attempting to derive the (very uncertain) best fit. We thus start with continuous disks with a single power law to define the flaring and density profile and depart from that model by adding a differentiated inner disk if the SED cannot be reproduced with a simpler model.
For each SED, several models were run, aiming at reproducing the observations with very different disk parameters in order to probe all possible scenarios. In most cases, the result is that each given SED may be equally (or nearly equally) well reproduced by very different models within a non-continuous parameter space. Very often, none of the models produces a good fit at all wavelengths, which is an indication that the disk structure is likely more complex than what our simple models consider.  Table \ref{models-table} contains a summary of the models, and below we provide an object-by-object discussion. The resulting models are presented in Fig.~\ref{models-fig}, together with the observed SED and the IRS/{\it Spitzer} spectra. Specific objects are discussed in the following sections.


\subsubsection{RECX-5}

The stellar parameters considered for RECX-5 are a stellar mass $M_*$ = 0.3 M$_\odot$ and a stellar radius $R_*$ = 0.7 R$_\odot$. The SED of RECX-5 is characterized by a very strong silicate feature \citep{Bouwmanetal2006,Sicilia-Aguilaretal2009}. It is now found to have also a very high submillimeter flux, which requires a total disk mass of about 1.8 $\times$ 10$^{\,-3}$ M$_\odot$ (or 6 $\times$ 10$^{\,-3}$ $M_*$) to reproduce the LABOCA emission with the given grain sizes, and a maximum scale height of 0.06 $\times$ $R_{\rm \,out}$. 
RECX-5 has been modeled in detail by \citet{Bouwmanetal2010}, who suggested that, in addition to a small inner hole 0.6 au in size, a second gap centered at 24 au was also necessary to reproduce the flux.
Details of this model, which was also created using RADMC-2D, can be found in the original paper.

The results of our modeling are that, while an essentially empty hole (filled only with small, optically thin dust grains including crystals) is required to reproduce the SED, the SED alone does not require including additional gaps at larger distances to reproduce the Herschel and LABOCA fluxes
(see Fig.~\ref{models-fig}). The required size of the inner hole is of the order of 0.5-1 AU, with a
temperature for the inner disk rim of about 200 K, but the fact that the disk SED can be well-reproduced using very different structures at moderate to large radii is a sign of the degeneracy resulting from unresolved observations.

Given the number of disks that are observed to have asymmetries and large-scale gaps \citep[e.g.,][among many others]{vandermarel13,garufi14,avenhaus14,alma15,isella16,benisty18} and that the relation between accretion and Herschel fluxes observed in transition disks hints toward the presence of further gaps \citep[hard to detect from the SED alone;][]{Sicilia-Aguilaretal2015}, we cannot exclude the presence of further radial and azimuthal asymmetries in the disk of RECX-5, but spatially resolved data will be required for final confirmation.

\subsubsection{RECX-9}

For this disk, the mismatch (or discontinuity) between the mid-IR spectrum and the far-IR data makes it very hard to produce a good fit without some drastic radial changes at intermediate distances (12-25 au). Having a gap and/or a change in flaring/settling at these distances can reproduce the observed profile. The location of this asymmetry is roughly consistent with the position of the known companion at $\sim$ 0.2 arcsecs \citep[$\sim$ 20 au;][]{Bouwmanetal2006}. There is still substantial submillimeter emission, suggesting that the disk is not fully truncated so there is still some circumbinary material present, which may account for the extra disk mass inferred from the submillimeter flux. We checked if the submillimeter emission could be contaminated by another source.  The {\it Planck} maps between 30 and 70 GHz do not reveal any nearby source
that could affect the submillimeter flux of RECX-9\footnote{The nearest sources are quasars at $\sim$6$^\circ$ and 8$^\circ$.}.  
The lack of a strong silicate feature is a sign of strong settling and/or a strong population of large grains in the inner disk, both characteristic of an evolved protoplanetary disk. In fact, the disk can be fitted with the same combination of small and large grains extending throughout the whole disk.
Some care has to be taken regarding the fit of the far-IR data, given that the mismatch between the Spitzer 70 $\mu$m and the Herschel/PACS data points may suggest either contamination by cloud emission or some episode of accretion variability, although there is no evidence that RECX-9 suffers accretion variations beyond a factor of few (see Table \ref{macc}).

\subsubsection{RECX-11}

Besides being relatively settled and rich in large grains (which causes the observed low-contrast silicate feature), the SED of RECX-11 does not display any clear evidence of a distinct radial substructure, except for maybe a slightly puffed-up inner disk and a potential change in flaring/gap at intermediate radii (see Fig.~\ref{models-fig}). The strong submillimeter emission is characteristic of a relatively settled (or flattened), but still massive disk. Note that the models here assume that grains of all sizes are well-mixed, which may not be the case, especially regarding the larger grains that dominate the far-IR and submillimeter points.
We also assume that there is no difference in composition between the inner and the outer disk.

\subsubsection{ECHA J0820}
One of the members of the $\eta$\,Cha extended population discovered by \citep{Murphyetal2010}, together with ECHA J0801, ECHA J0820 also shares with the former a very low near-IR emission. The disk has substantial emission at 22 $\mu$m as well as at 870 $\mu$m setting strong constraints on the presence of gaps and holes in the inner disk. In this case, a gap or a partly evacuated hole 2-3 au in size offers a very good fit to the SED. Because of the very low near-IR emission, the inner disk is required to be both depleted and flattened, compared to the outer disk.

\subsubsection{ECHA J0843}
Also known as RECX-15 and ET Cha, fitting its strong silicate feature and the high emission from 18 $\mu$m on requires a change of flaring angle, composition, or slope in the innermost disk, which could be a sign of a gap in the disk at small radii ($\leq$ 0.5 au). The disk is consistent with a
strong and optically thick inner part, followed by a gap or a change in vertical scale height. The precise parameters of the gap depend strongly on the dust properties and on the shape of the inner rims of the gap, which we are keeping as a simple vertical slope in this case. Since the silicate feature is strong, we use a small-grain (amorphous silicate and forsterite, 0.1 $\mu$m in size) dominated inner disk and an outer disk with the usual larger grain distribution, but a similar result can be attained by including the small grains on the outermost layers of the disk or within the gap.

\subsubsection{ECHA J0841}
This disk presents an anomalously high submillimeter datapoint, compared to the rest of the SED, but since it is a 2.9\,$\sigma$ (marginal) detection, it may be considered as a possible upper limit. The innermost disk SED, including the Herschel data, is consistent with a strongly settled system with very low mass, while the mass predicted from the submillimeter flux, despite its uncertainty, is higher than what one would expect for a standard, continuous disk with such a flat inner disk. One possibility would be the presence of large gaps at large radii (to explore this possibility, we would need data in the 200-700 $\mu$m range), or a nearly geometrically flat disk (which could be caused by settling in a large-grain, gas-poor disk. Another possibility would be that the millimeter flux is overestimated and thus an upper limit. Examination of the IR images does not reveal any evidence of a potential contaminant submillimeter source at distances relevant to the APEX beam. In any case, the existing near- to far-IR data show a very evolved, settled system.

\subsubsection{ECHA J0844}

The structure of this disk is very similar to ECHA J0843, although the gap/wall is located at larger distances and the disk appears more settled and evolved. We again encounter the same difficulties, that the flux in the mid/far-IR depends strongly on the properties of the gap and the gap walls, leading to large uncertainty in a broad and non-continuous parameter space. None of the simple models used here offers a perfect fit.
The comparison with the well-mixed disk (where dust follows the hydrostatic equilibrium distribution of the gas) suggests that the disk is artificially puffed up in the innermost part or at least, that it has a puffed-up inner rim. Varying the dust distributions and surface densities in the inner disk could be done to improve the fit, but the uncertainty cannot be resolved without further (if possible, spatially-resolved) information.

\subsubsection{$\eta$\,Cha}
\label{etacha-sec}
\begin{figure}
\centering
\includegraphics[width=0.99\linewidth]{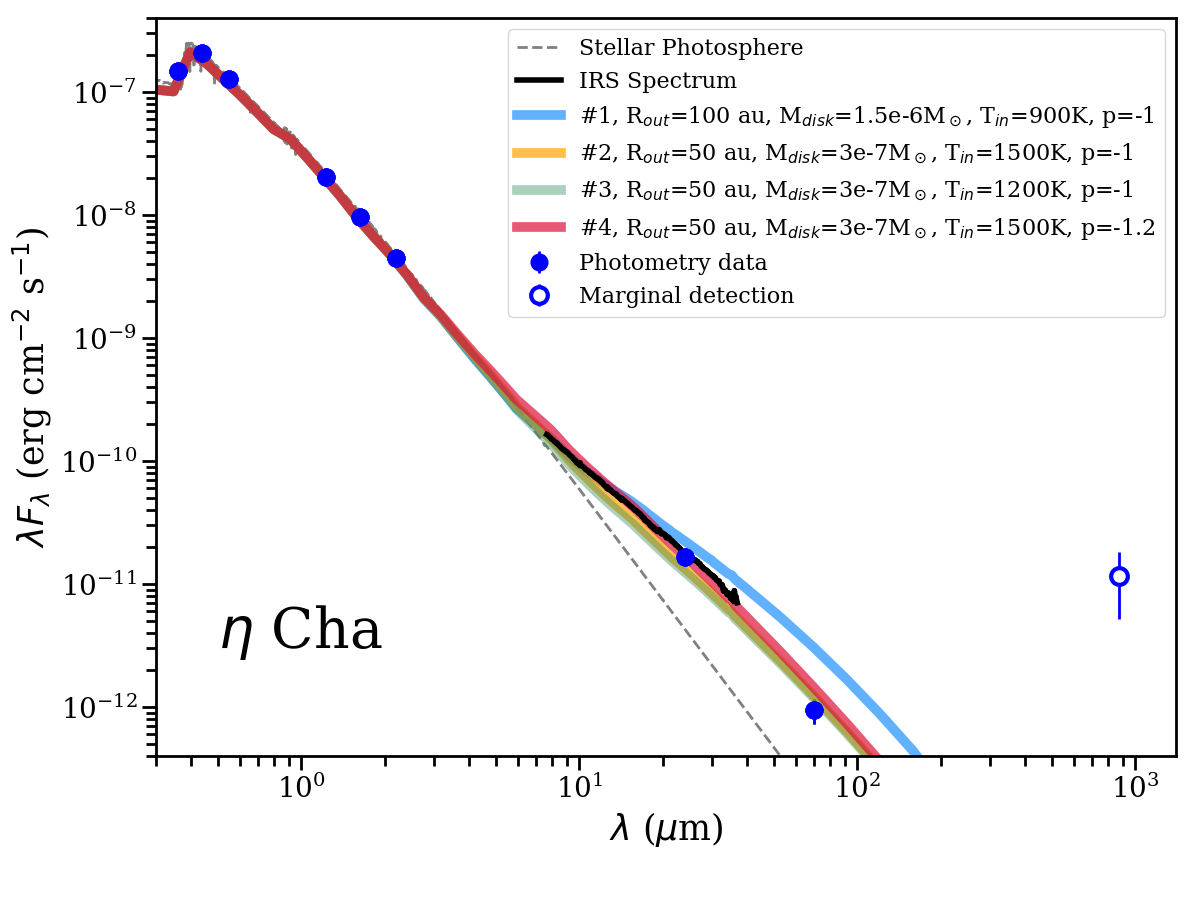}
\caption{SED fit for $\eta$\,Cha. Note that the LABOCA 870 $\mu$m point is not fitted, as it is most likely contaminated as explained in the text.}\label{etacha_sed_fit}
\end{figure}

The intermediate-mass star $\eta$\,Cha is a special case, compared to the rest of the objects, since it has a disk that seems to be intermediate between a protoplanetary disk and a debris disk, but it does not seem to have a large inner hole or gap and the disk may start straight away at the dust destruction radius. The disk had been also proposed to arise from decretion, as in a Classical Be star, but optical spectroscopy does not reveal the required gas features \citep{Sicilia-Aguilaretal2009}. We have further examined in detail the existing spectroscopy of $\eta$\,Cha in the public ESO database, and none of the observations 
reveal the gas lines expected in a Classical Be star, so that the disk around $\eta$\,Cha appears to be either an evolved protoplanetary disk or a young debris disk extended over a relatively large range of radii. 

When fitting the SED for $\eta$\,Cha, we notice that the LABOCA marginal detection must be either a non-detection, contaminated or the error underestimated, because it is not possible to reproduce it with any sensible model of disk associated with $\eta$\,Cha. The emission would need to arise from very cold dust ($<$10 K) and therefore, be located at a very large distance from the relatively luminous $\eta$\,Cha. The very low temperature would require a very large dust mass (several Jupiter masses in dust alone) so that it likely corresponds to molecular cloud emission (or it is rather a non-detection than a marginal detection).

The inner disk requires optically thin dust, with a minimum size of the order of 20 $\mu$m to avoid silicate emission. As mentioned, an inner hole is not needed, but the low flux at 70um suggests that the disk is relatively small ($<$100 au, with the best fit being around 50 au). The flaring of the disk cannot be constrained because it is optically thin, but the fact that it is very hard to reproduce the shorter and longer wavelengths with a uniform disk suggests that the disk matter may be distributed asymmetrically in the inner vs. the outer disk, or that the disk has a slightly faster-falling power law of the surface density with radius (the best model has an exponent of -1.2, instead of -1 which is used in the rest of cases). The results of the fitting procedure are shown in Fig.~\ref{etacha_sed_fit}.

\begin{figure*}[h!]
\centering
\begin{tabular}{cc}
\includegraphics[width=0.38\linewidth]{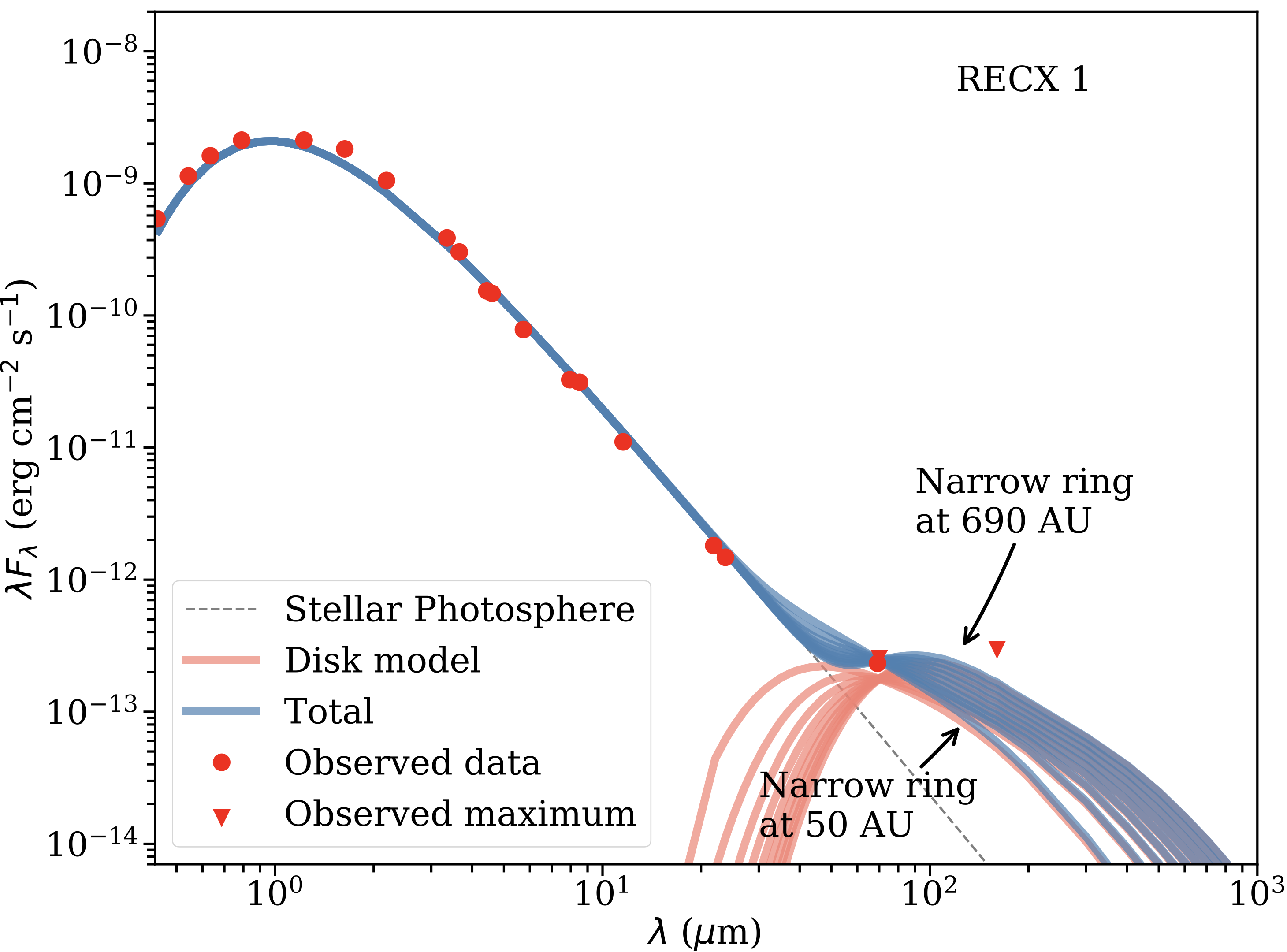} &
\includegraphics[width=0.38\linewidth]{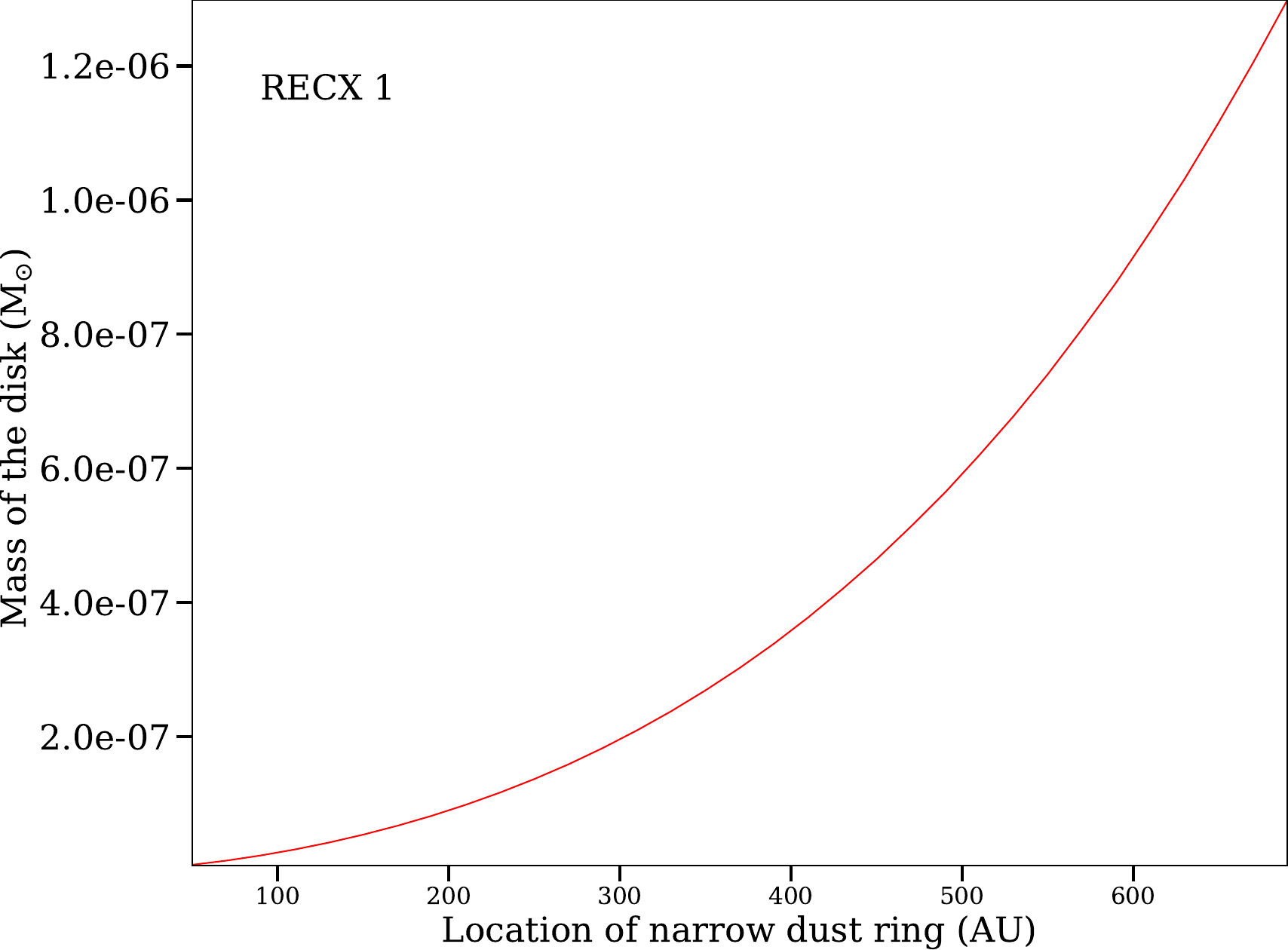} \\
\includegraphics[width=0.38\linewidth]{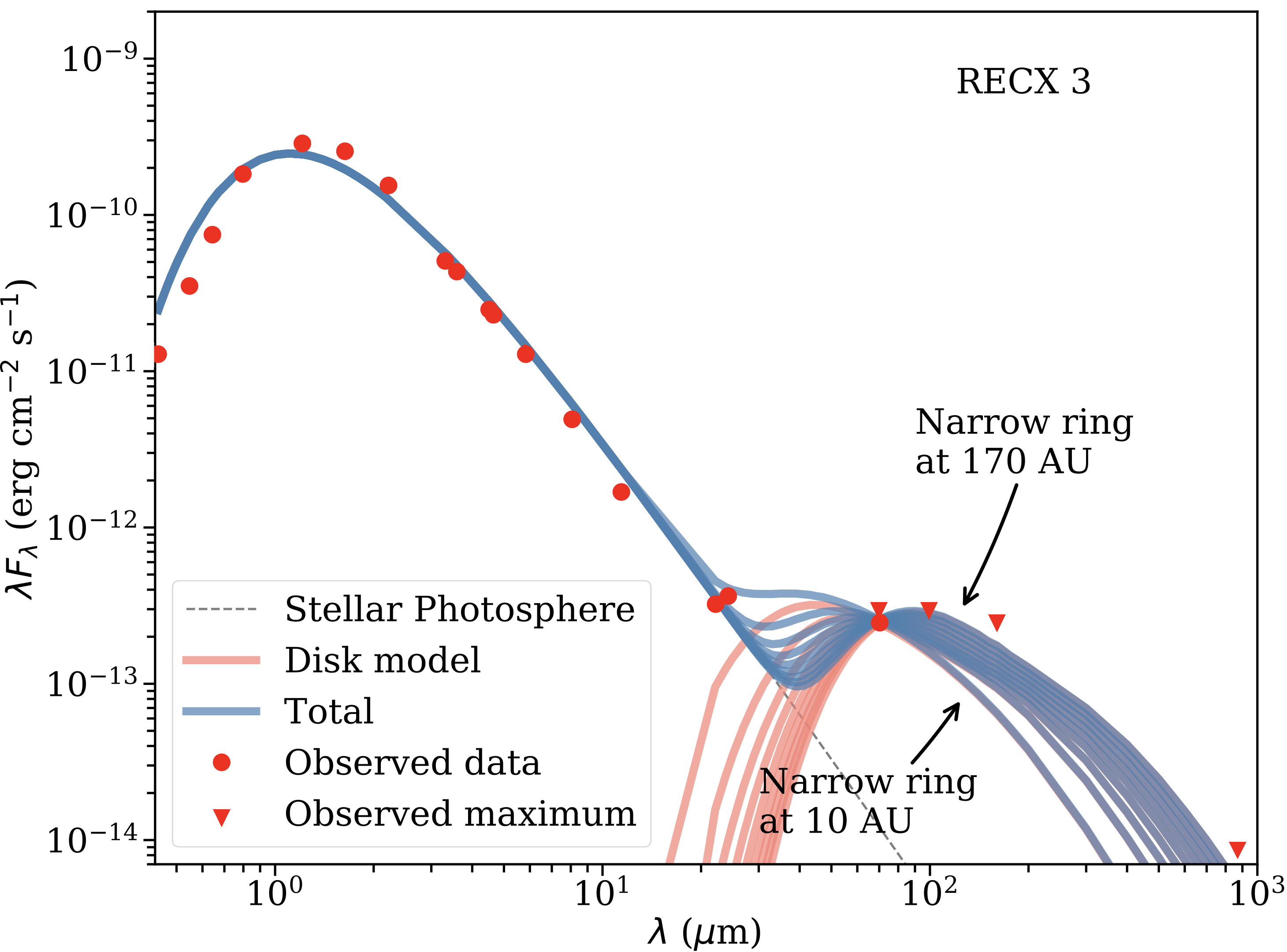} &
\includegraphics[width=0.38\linewidth]{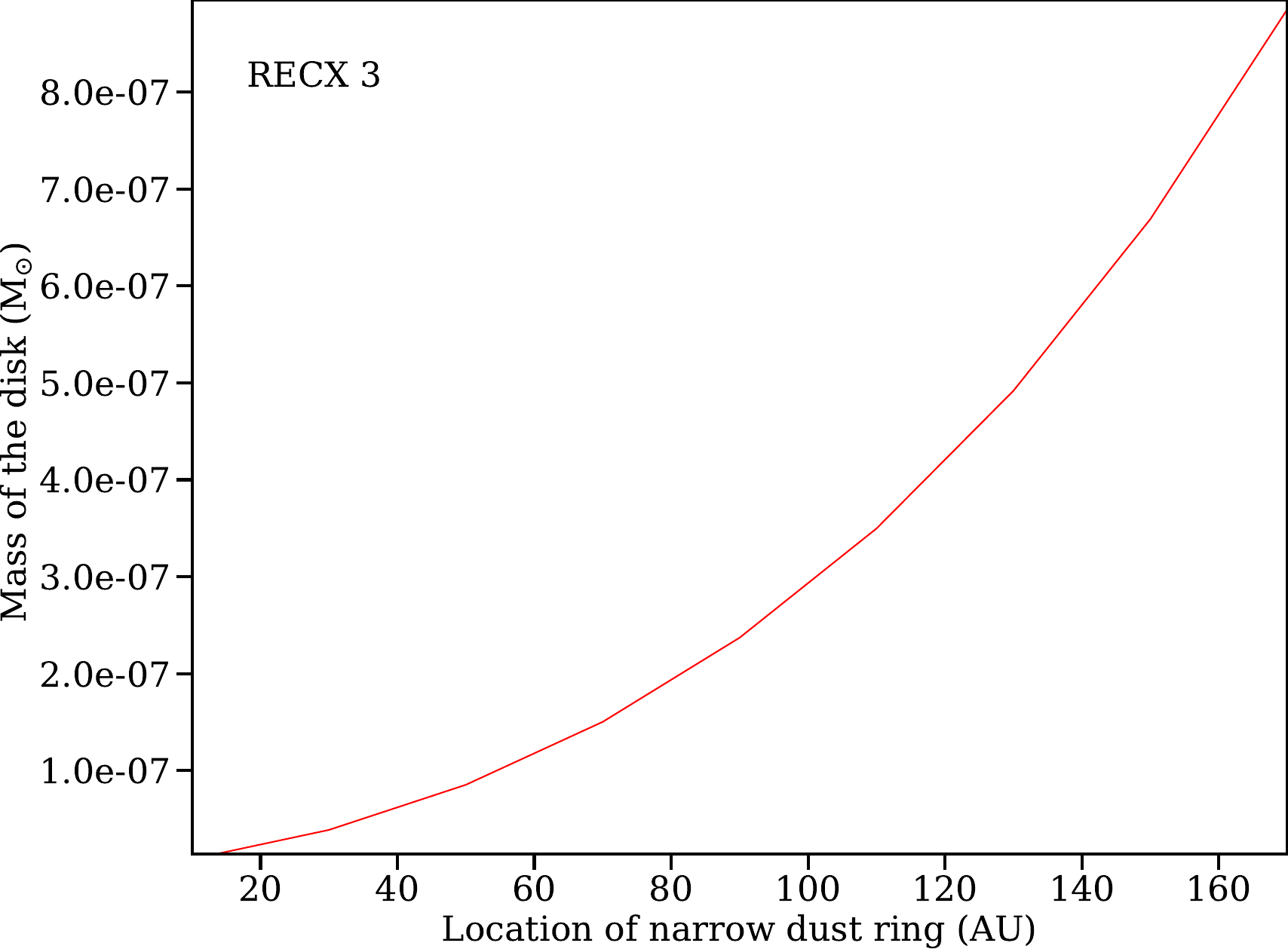} \\
\includegraphics[width=0.38\linewidth]{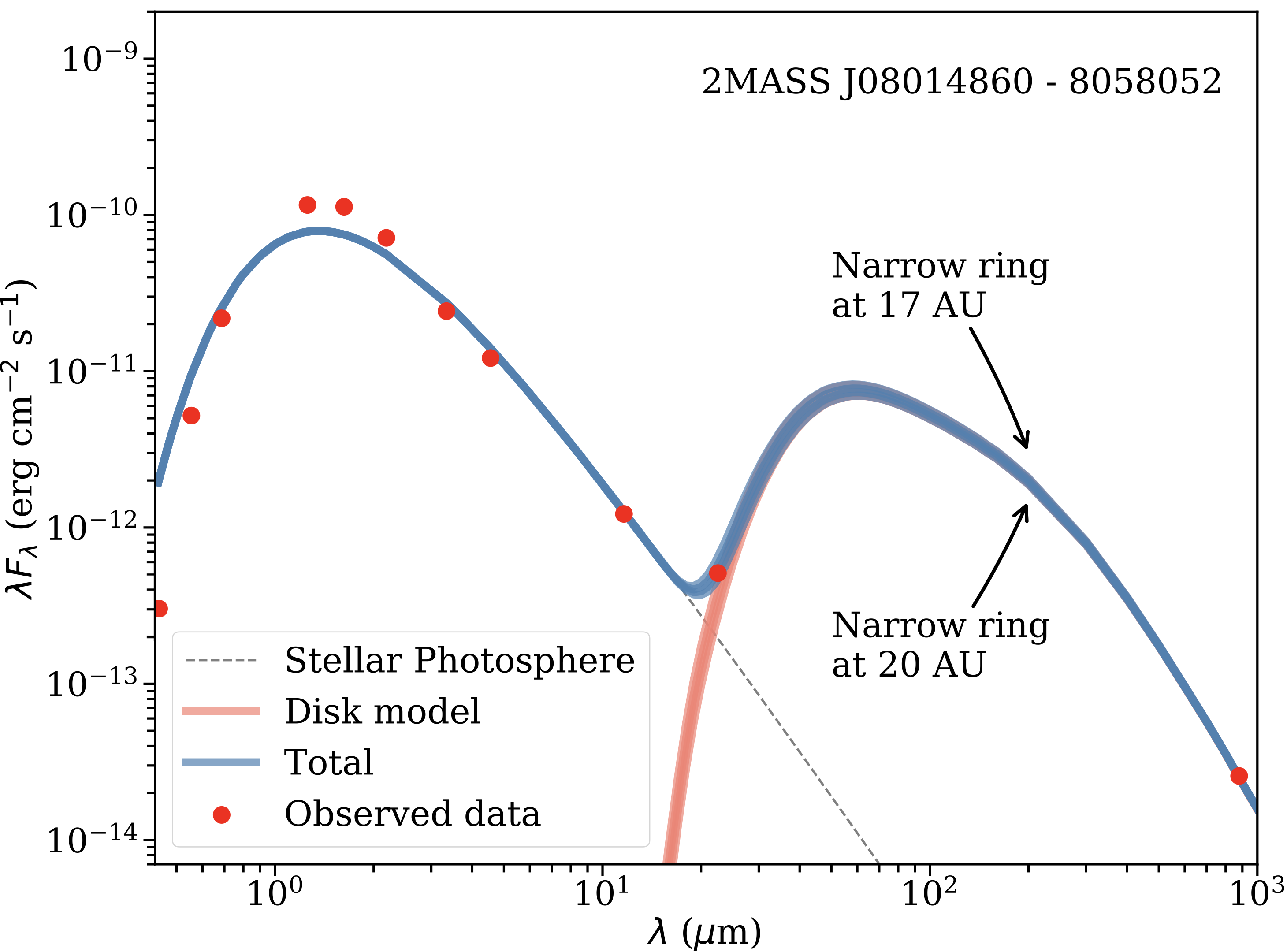} &
\includegraphics[width=0.38\linewidth]{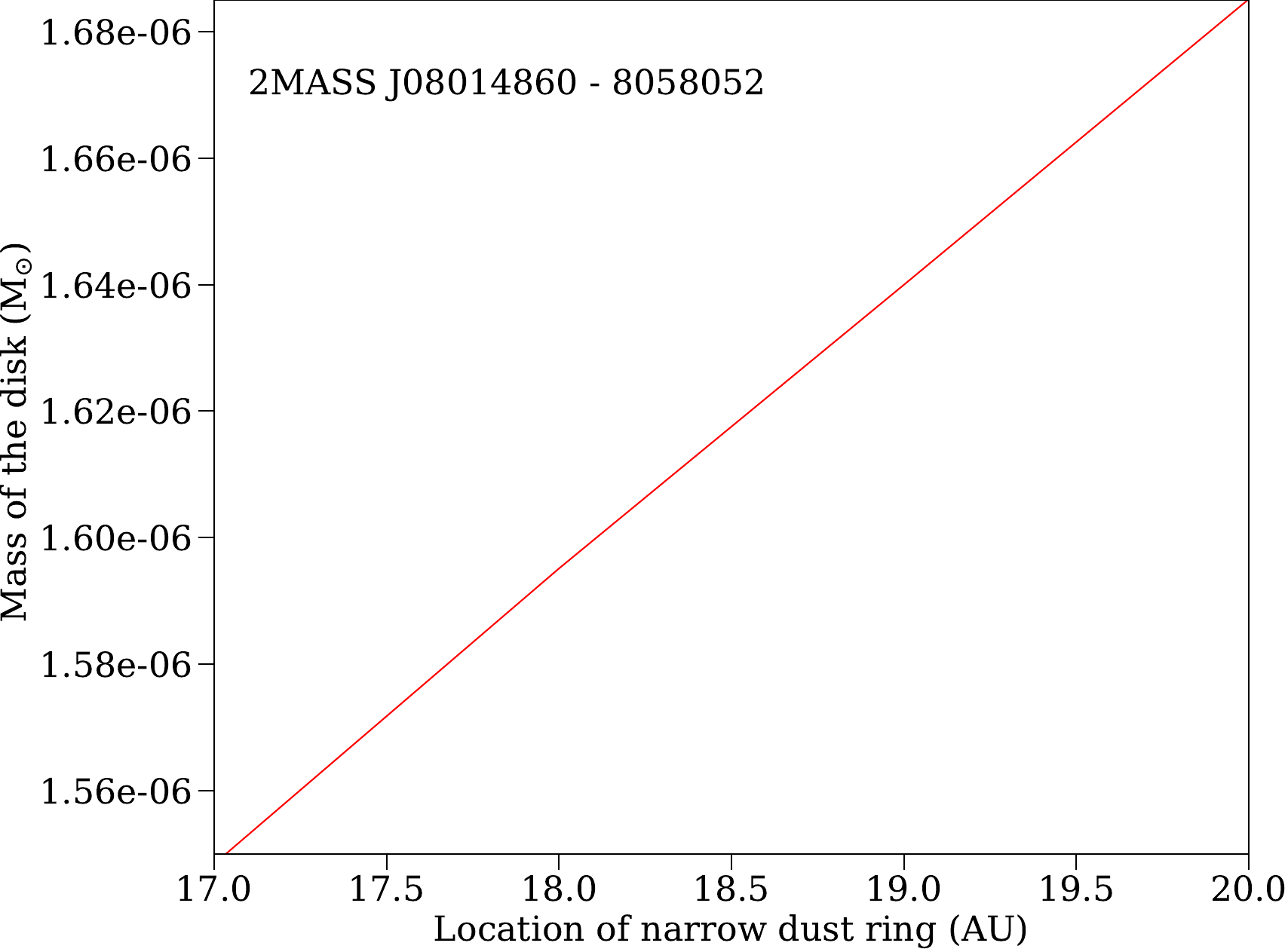}\\
\includegraphics[width=0.38\linewidth]{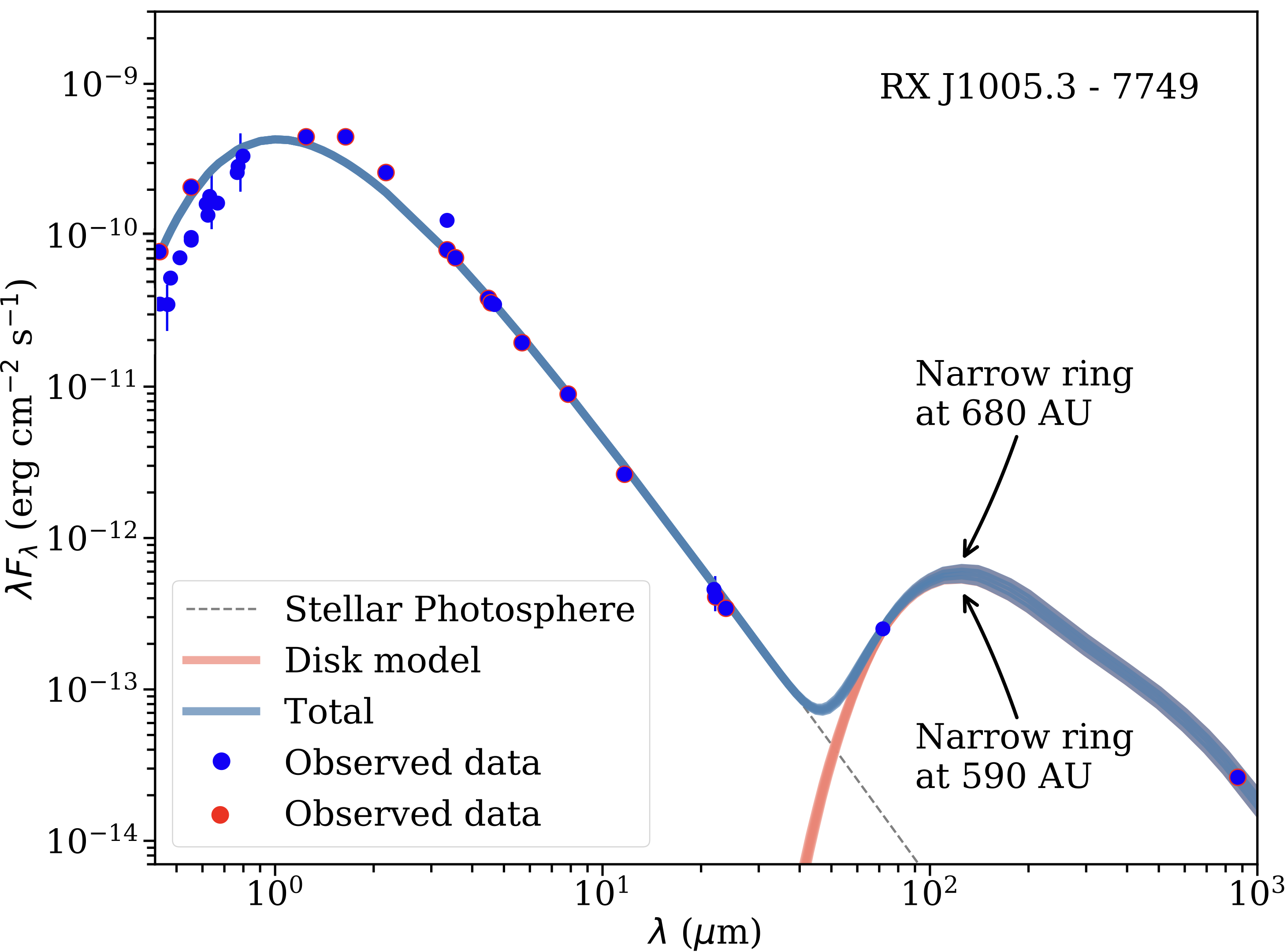} &
\includegraphics[width=0.38\linewidth]{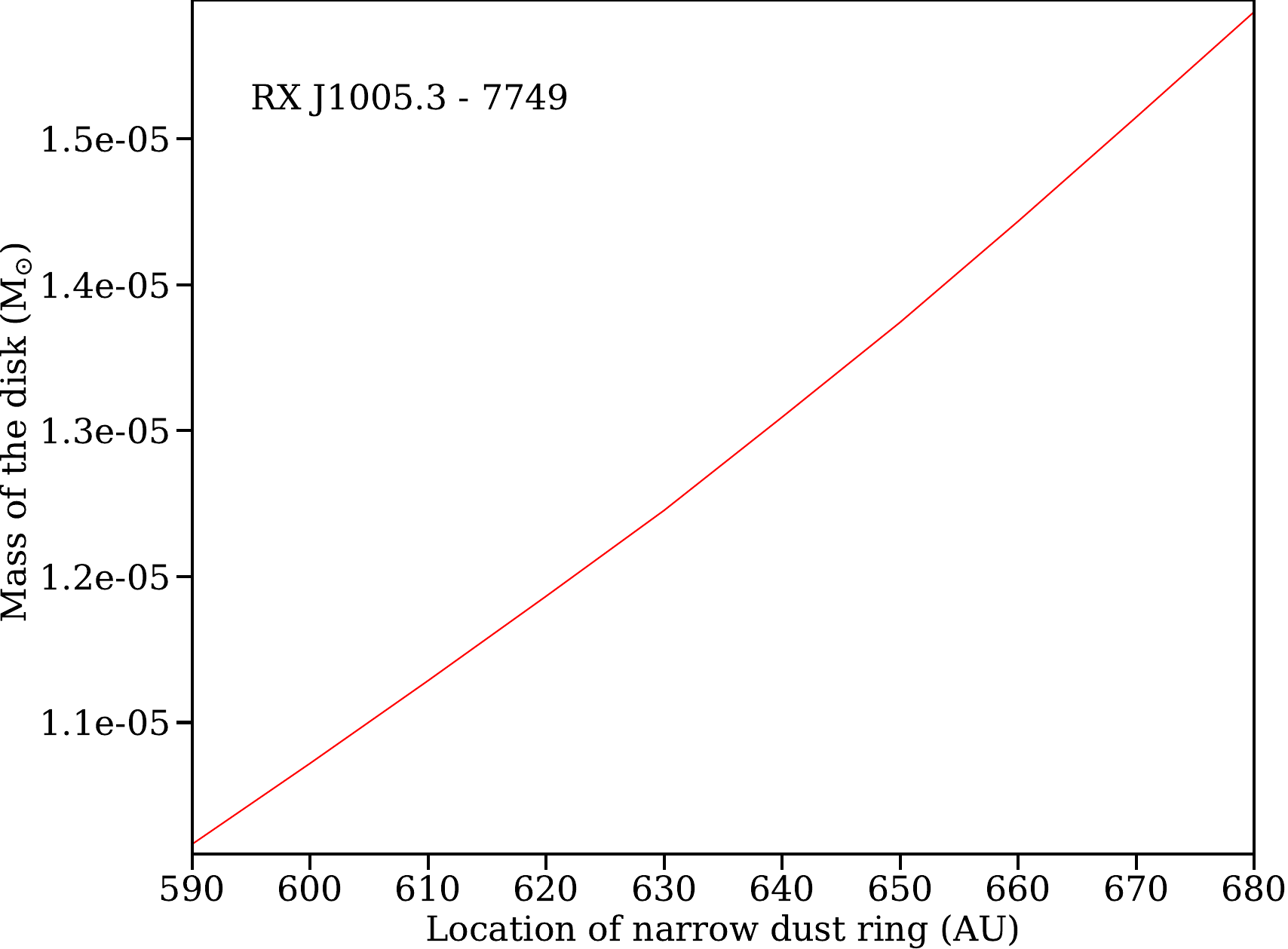}\\
\end{tabular}
\caption{{\it Left:} SED models for the debris disks. The photometry data are marked by red circles and triangles (the upper limits).  Stellar photospheric, disk, and total emission (stellar photospheric + disk models) are represented by grey dotted lines, and red and blue continuous lines, respectively (see Table \ref{models-table} for details on individual models). The blue dots in the SED of RX J1005.3-7749 show the detection from \citet{Murphyetal2010}. {\it Right:} Results of the SED modeling of the debris disks showing the mass of the disk as a function of the location of the narrow dust ring.}\label{models-dd}
\end{figure*}

\subsection{Modelling the debris disks with DMS}

We define Class III sources with ${f_{\rm\,dust}\, (\equiv L_{\rm\,dust}/L_{\rm\,*})\,<\, 10^{\,-2}}$ as debris disks. These disks are modeled assuming an optically thin emission from a dust distribution around a stellar source. 
As the first approximation, the debris dust in the disk is distributed in a very narrow ring with a fixed radial width of 20 au  (Fig.~\ref{figVibStab}). The re-emission of the stellar radiation from the disks is reproduced by varying the location of the inner and outer ring radii between 10 and 700 AU, respectively, as well as the total disk mass. This approximation of the ring emission is well supported by recent observations at different wavelengths of resolved debris disks distributed as a narrow ring {\citep[e.g., ][]{Kalasetal2013, Millietal2017, Schneideretal2018, Marino2018, Booth2023}. 
Debris dust material is assumed to be astrosilicates with a bulk density of 3.5 g cm$^{-3}$ \citep{Draine2003}. We note that most of the debris disks or Class III (without debris material around the central star) objects are stars with very late spectral type. For these cases, the radiation pressure exerted by these stars is insufficient to achieve a blowout size, particularly for stars with temperatures below 5250 K (\citealp{Kirchschlager2013}). Thus, we consider a range of dust grain sizes from 0.1 $\mu$m to 1000 $\mu$m}, where the grain size distribution $n\,(a)$ follows the power-law distribution $n\,(a)$ $\propto$ $a^{\rm - 3.5}$ 
\citep{Dohnanyi1969}. 

We use a newly developed software tool \textit{DMS} \citep[Debris disks around Main sequence Stars;][]{Kimetal2018} which is optimized for the simulation of debris disks, i.e., optically thin systems. In particular, it allows us to simulate scattered light and thermal dust reemission images, the continuum spectral energy distributions, and scattered light polarization images. The optical properties of the dust grains are computed using the tool \textit{miex} \citep{Wolfetal2004}. The characteristics of the central stars are listed in Table~\ref{fluxes}.

The results of the modeling (Fig.~\ref{models-dd}) of the debris disks show that in two cases the ring the close to the central star. The first case is represented by 2MASS J08014860,  discussed in detail in Sect.~\ref{j0801}, while the second case is represented by RECX 3, an M3 star, where the infrared photometry already suggested the presence of cold material around the star. Our modeling allows us to define a range in distance between 20 and 160 au for a ring with a mass of 1-9\,$\times$\,10$^{-7} $M$_\odot$. In the case of RECX 1, a K7 star, where two detections in the far-infrared lie above the stellar photosphere, the missing LABOCA observation shows the poor fitting of the SED which leads to a poor constraint of the location of a 0.2-1.2\,$\times$\,10$^{-7} $M$_\odot$ disk between 100 and 700 au. Finally, the M1 object, RX J1005.3-7749\footnote{Gaia DR2 5202670052321415040}, was first classified as a  pre-main-sequence object by \citet{Covinoetal1997} and then \citet{Murphyetal2010} as a possible dynamical $halo$ member. It harbors the most massive ring (1.0-1.5\,$\times$\,10$^{-5}$ M$_\odot$) at a location between 590 and 680 au, more distant than expected, meaning that the dust temperature is extremely low. This, in turn, results in a remarkably high derived dust mass. 
We note that there is a possibility that the millimeter emission attributed to the debris disk could be contaminated by background galaxies or asymptotic giant branch stars. Similar issues have been addressed in the previous observations, in particular, for the Class III young stellar objects, showing similar infrared excesses that lead to similar expected signatures in observations (e.g., \citealp{Oliveira_2009, Evans2009}). However, with the recent release of the second set of data from the Gaia mission \citealp{Gaia1_2018} and the availability of high-resolution optical spectra in the present study, allowing for more accurate identification and classification, we are better equipped to differentiate and exclude these sources from consideration in our analysis.

Tentative SED models have been also carried out for the probable debris disks, RECX 6 and RECX8. We explore different disk models, represented by blue lines in Fig.~\ref{models-ddbad} in the Appendix, but none of them were able to reproduce all of the observed data, as is the case with other systems. We note that RECX 8 (RS Cha) is an eclipsing binary system with two components of similar mass and luminosity, which is confirmed using the TESS lightcurve by \citep{Steindl+2021}. Furthermore, the third low-mass component has been recently proposed by \citet{Woollands+2013}. Due to the complexity of this system, it cannot be approximated by our assumed single thin-ring model. 

\begin{sidewaystable*}
\caption{Parameters used in the disk models. Note that R$_{\rm gap}$ and T$_{\rm gap}$ refer to the location of a radial discontinuity, which may involve not only a thin gap but a change in radial properties (e.g., dust grain properties, vertical scale height, density power law). Note that for some models, we set the inner disk temperature in RADMC, and the distance is calculated accordingly, although the opposite is also feasible. In this table, we list the distances as well because they are easier to compare with resolved observations and other models in the literature, and the derived quantities are marked with~$^{*}$. The parameter p$_{H/R}$ corresponds to the exponent in ($H/R$) $\propto$ R$^{\,\rm{p}}$, and p$_{\Sigma\,\rm{dust, in}}$ is the power law vs radius for the surface density in the inner disk.
The notes include any further deviation from the standard model (see text) not reflected in the previous parameters.} 
\label{models-table}      
\centering               
\begin{tabular}{r l c c c c c c c c c c c l}  
\hline\hline                
\noalign{\smallskip}
Source & Model& T$_{\rm eff}$ & R$_{*}$		&R$_{\rm in}$ 	& T$_{\rm in}$ & R$_{\rm gap}$ & T$_{\rm gap}$ & H/R$_{\rm out}$ & H/R$_{\rm in}$ &   $\Sigma_{\rm dust,in}$ &p$_{\rm H/R}$ & M$_{\rm disk}$  & Notes \\
 	& \# 	& (K) &R$_\odot$& (au)     	&  (K)     &  (au)   &  (K)  & (out/min)   &  (out/min)  & (g cm$^{-2}$) &   (in, out) & (M$_\odot$)  &  \\
\hline
\noalign{\smallskip}
RECX-5 	& \#1 & 3350&0.76&0.018 & 1480$^*$	& 1$^*$ 	& 200	& 0.06/0.005 & 0.04/0.01 & 1e-5 & 1/7, 1/7& 1.8e-3 & Schmidt Nr=10, p$_{\Sigma \rm dust,in}$=-1.4,\\
  &&&&&&&&&&&&& Small grains\\
RECX-9 	& \#1 & 3305&0.95&0.02$^*$	& 1500  & -- 	& -- 	& 0.07/0.005 & --  	& --   & 1/7	& 1.4e-4 & \\
	& \#2 &&& 0.07 	& 830$^*$	& 12 	& 63	& 0.1/0.005  & 0.3/0.05 & 1e-5 & 1/7, 1/7& 1.2e-4 & \\
  	& \#3 &&& 0.07 	& 830$^*$	& 20$^*$ 	& 50	& 0.2/0.005  & 0.2/0.05 & 1e-5 & 1/7, 1/7& 1.4e-4 & \\
RECX-11 & \#1&& 	& 0.05  & 1500$^*$	& 0.64  & 430$^*$	& 0.08 & 0.12 	& 2e-5 & 1/7, 1/7& 2.5e-4 & \\
  	& \#2  &&& 0.05  & 1500$^*$	& 0.64  & 430$^*$	& 0.05 & 0.14 	& 5e-5 & 1/7, 1/7& 3.4e-4 & \\
  	& \#3  &&& 0.05  & 1500$^*$	& 0.64  & 430$^*$	& 0.05 & 0.14 	& 5e-5 & 1/7, 1/7& 3.4e-4 & Small grains in inner disk\\
  	& \#4 	&&& 0.05  & 1500$^*$	& 0.64  & 430$^*$	& --   & -- 	& 5e-5 & -- & 3.4e-4 & Well-mixed gas and dust\\
J0801  	& \#1&3280& 0.62	& 0.02 & 1250$^*$  & 12.5	& 50$^*$	& 0.03 & 0.03 	& 1e-5 & 1/7, 1/7& 1e-4 & \\
  	& \#2  & &&0.02 & 1250$^*$  & 12.5  & 50$^*$	& 0.03 & 0.03 	& 1e-6 & 1/7, 1/7& 8e-4 & \\
 	& \#3  & &&0.02 &  1250$^*$	& 19.5  & 40$^*$	& 0.02 & 0.03 	& 1e-6 & 1/7, 1/7& 8e-4 & \\
  	& \#4  & &&0.02 &  1250$^*$	& 19.5  & 40$^*$	& 0.01 & 0.01 	& 1e-6 & 1/7, 1/7& 8e-4 & \\
J0820  	& \#1 &3280&0.60	& 0.03 	& 1000$^*$	& 3 	& 100$^*$	& 0.17 & 0.05 	& 1e-4 & 1/7, 1/7& 4e-4 &  \\
  	& \#2 	&&& 0.03 	& 1000$^*$	& 2.1 	& 120$^*$	& 0.17 & 0.05 	& 1e-4 & 1/7, 1/7& 4e-4 & \\
  	& \#3 	&&& 0.03 	& 1000$^*$	& 3 	& 100$^*$	& 0.05 & 0.20 	& 1e-4 & 1/7, 1/7& 4e-4 & \\
J0841  	& \#1 && & 0.03 	& 900$^*$	& 0.1$^*$	& 495 	& 0.04/0.005 & 0.1/0.05 	& 1e-5 & 1/7, 1/7& 1.2e-4 &  \\
 	& \#2  &&& 0.03 	& 900$^*$	& 0.1$^*$	& 495	& 0.03/0.005 & 0.1/0.05 & 1e-5 & 1/7, 1/7& 4e-5 &    \\
  	& \#3 &&& 0.03 	& 900$^*$	& 0.1$^*$	& 495  	& 0.03/0.005 & 0.1/0.05  & 1e-5 & 1/7, 0	& 8e-5 	&   \\
	& \#4  &&& 0.03 	& 900$^*$	& 1.5$^*$	& 130  	& 0.02/0.005 & 0.1/0.01 & 3e-5 &  0, 0	& 5e-4 	&   \\
J0843  	& \#1 &3430& 0.79& 0.03 	& 1200$^*$	& 0.18$^*$	& 495 	& 0.28 & 0.17 	& 5e-3 & 1/7, 1/7& 3e-4 &  Small grains in inner disk\\
  	& \#2 &&& 0.03  & 1200$^*$	& 0.9$^*$	& 220 	& 0.25 & 0.17 	& 5e-3 & 1/7, 1/7& 2.4e-4 &  Small grains in inner disk \\
J0844  	& \#1 &3145&0.37& 0.03 	& 750$^*$	& 3$^*$ 	& 78	& -- & -- 	& 1e-4 & --	& 3.9e-4 & Well-mixed gas and dust \\
  	& \#2 	&&& 0.03 	& 750$^*$	& 2$^*$  	& 92	& 0.15 & 0.2 	& 1e-4 & 1/7, 0	& 3.9e-4 &  \\
  	& \#3 	&&& 0.03 	& 750$^*$	& 3$^*$ 	& 78	& 0.17 & 0.1 	& 1e-4 & 1/7, 0 & 3.4e-4 &  \\
  	& \#4 	&&& 0.03 	& 750$^*$	& 3$^*$ 	& 78	& 0.15 & 0.1 	& 1e-4 & 1/7, 0 & 3.9e-4 & \\
$\eta$\,Cha   & \#1&10500&2 & 1.6$^*$ & 900  & --  & -- & 0.01/0.002 & -- &  -- & 0 & 1.5e-6 & Large grains, not flared, 100 au \\
 & \#2 &&& 0.6$^*$  & 1500 & -- & -- & 0.06/0.002 & -- &  -- & 1/7 & 3e-7  &  Large grains \\
  &  \#3 &&& 0.9$^*$ & 1200 & -- & -- & 0.03/0.002 & --  & -- & 1/7 & 3e-7 &   Large grains \\
   & \#4 &&& 0.6$^*$ & 1500 & -- & -- & 0.06/0.002 & -- & -- & 1/7 & 3e-7 & Large grains, 50 au. Surface density\\
  &&&&&&&&&&&&& power law p$_{\Sigma \rm dust}$=-1.2 \\
\hline                        
\noalign{\smallskip}
\end{tabular}
\end{sidewaystable*}
\subsection{ECHA J0801}
\label{j0801}

ECHA J0801 (2MASS J08014860) is one of the M5 sources classified by \citet{Murphyetal2010} as possible halo members of $\eta$\,Cha, and it may be a binary system due to its position in the color-magnitude diagram and the different high-velocity measurements \citep{Murphyetal2010}. Dynamical simulations suggested the system was ejected from the cluster. \\
Although ECHA J0801 has a negligible excess only in the mid-IR, we find a significant submillimeter flux, which might suggest a massive disk. The presence of substantial circumstellar material can also explain the observed variable accretion \citep{Murphyetal2011}. 
We highlight, however, that this high submillimeter flux can be contaminated by the local diffuse nebulosity detected by {\it Planck} between 30 and 217 GHz. Taking into account this probable contamination, the submillimeter detection will be considered as an upper limit for disk emission. However, since the nature of such a system is still under debate, we model the SED using both RADMC 2D and DMS. 

Since the stellar photosphere had not been modeled previously, we use the stellar photosphere of ECHA J0841 (with a slightly different radius to account for the differences in luminosity) to simulate the stellar emission, given that the two stars are very similar. Using RADMC 2D, the lack of far-IR observations does not allow us to pinpoint the disk structure at intermediate radial distances, but the object is consistent with a nearly depleted inner disk together with a rather massive outer disk, starting some point after 12 au.
Given its very low mass, it belongs to the class of transitional objects with very large gaps, which is rare around late-type stars.\\
Using instead DMS, ECHA J0841 is assumed to be a debris disk with an optically thin dust emission. With the same approximation as in the other debris systems, we find that ECHA J0841 is the only source with a massive narrow ring of 1.6-1.7 $\times$ 10$^{-6}$M$_\odot$ at between 17 and 20 au from the central star. \\
It is very interesting to highlight that both approaches led to a large gap of about 15 au and a very massive disk compared to other systems (the absolute disk mass is not comparable since the grain distribution and disk size are different).  

\subsection{The age of the $\eta$\,Cha association}

\begin{figure}
\centering
\begin{tabular}{c}
\includegraphics[width=8.5cm]{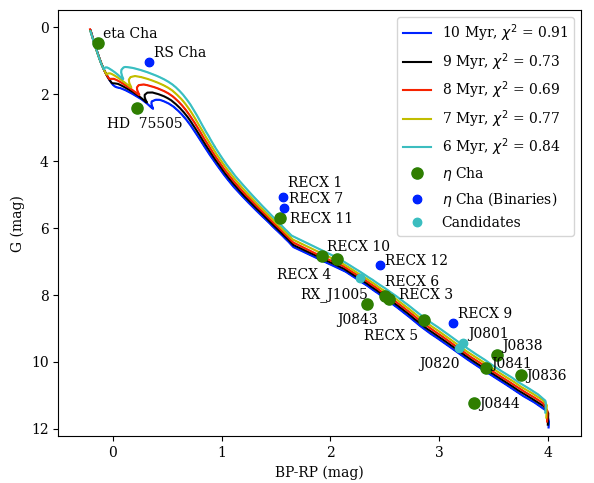}\\
\includegraphics[width=9cm]{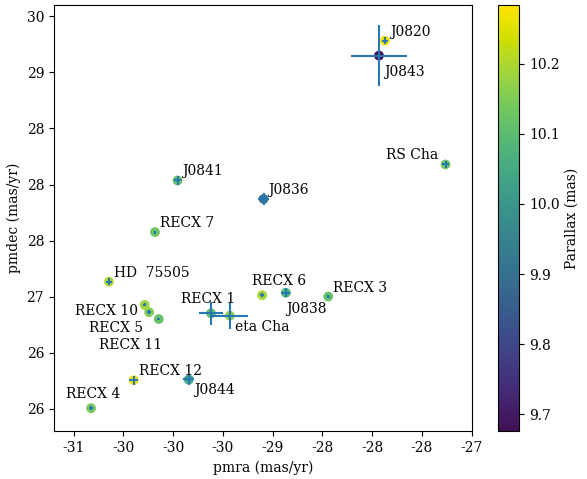}\\
\end{tabular}
\caption{\textit{Top:} Colour-Magnitude diagram of the $\eta$\,Cha association members from Gaia eDR3 photometry. Isochrones are overplotted and determined $\chi^2$ values are shown in the legend, each with $p$ values > 9e6. Confirmed binaries, as indicated in Tab.\,\ref{macc} are not included in the $\chi^2$ calculations. \textit{Bottom:} Proper motion plot from Gaia eDR3 astrometry. Markers are colored by the star's parallax, as shown by the color bar. RX J1005 (-38,15) and J0801 (-19,29) are not shown. \label{etacha_cmd}}
\end{figure}

Recent work on Gaia observations of the nearby associations \citep{gagneetal2018XII} reported an age of 11 $\pm$ 3~Myr for the $\eta$\,Cha association. This value was not based on a new analysis of the available Gaia DR2 data presented in that paper, but instead on the previous work of \citet{Belletal2015}. They compiled the M$_{\rm V}$, V-J color-magnitude diagram (CMD) of 18 cluster members, applying a weighted average of a few stars within the cluster with trigonometric parallaxes from the revised Hipparcos reduction of \citet{vanleeuwen2007}, giving a distance of 94.27 $\pm$ 1.18 pc. The final given age is the average from the fitting of four different sets of isochrones \citep{Dotter+2008, Tognelli+2011, Bressan+2012, Baraffe+2015}. 

We have used the Gaia eDR3 photometric fluxes and parallaxes \citep{2021A&A...649A...1G}, finding the best age for the association to be 8 $\pm $ 1\,Myr (Fig.\,\ref{etacha_cmd}). The mean parallax for the association from eDR3 is 10.11 $\pm$ 0.13 mas, resulting in a distance of $\sim$ 99.0 $\pm$ 1.3 pc. 
Absolute G band magnitudes were calculated using the individual parallaxes for 17 of the cluster members and the three halo members. The only star that does not have a parallax measurement in eDR3 is RECX-9, for which we use the average parallax. PARSEC \citep{Bressan+2012} and updated COLIBRI \citep{2020MNRAS.498.3283P} isochrones were over-plotted to the CMD, with a standard $\chi^2$ calculation giving the best fitting age of 8\,Myr. All $\chi^2$ calculations had a $p$-value > 9e6. However, due to the intrinsic variability of PMS stars and the corresponding expected to scatter, 7 and 9\,Myr are also likely with their respective low $\chi^2$ scores. The five confirmed binaries in the association (as indicated in Tab.\,\ref{macc}) were not included in the $\chi^2$ calculations. Each of these binary members is shown to lie above the youngest isochrone in Fig.\,\ref{etacha_cmd}. \citet{2007A&A...473..181A} calculate the age of the eclipsing binary RS Cha to be 9.13 $\pm$ 0.12\,Myr, which is in good agreement with our estimate for the rest of the association. J0838 and J0836 are also slightly above the youngest isochrone. 

Two members are positioned below the oldest isochrones, J0843 and J0844. As discussed, these two stars have extended disks with similar properties. Intrinsic variability of the stars or occultations from disk material may be the reason for their G band flux being lower than expected for the age of the association. Another possible explanation could be accretion, which strongly affects the spectra of both sources \citep[Fig.~7 in ][]{Rugeletal2018}, or scattered light from the large disks, which can make objects appear bluer than expected. The three halo members, which are further away from the cluster center, have similar ages to the rest of the association. J0820 is located in the same proper motion space as the association, however, RX J1005 and J0801 are outliers in the proper motion plot.

\section{Discussion}
\label{discussion}

\subsection{Disk properties}

Models of the SEDs of protoplanetary and debris disk systems in $\eta$\,Cha 
allow us to characterize the disk geometry and dust distribution. 
In particular, none of the observed disks can be well-fitted using a non-evolved disk model, with well-mixed dust and gas (no settling) and a smooth distribution of dust across the disk. The SEDs of all of the disks in $\eta$\,Cha show clear deviations from a standard protoplanetary disk model, such as radial variations of the dust properties, settling, mass depletion, and the presence of radial asymmetries or substructures(holes, gaps, and/or changes in the vertical scale height) in the disk.

As expected, individual disk models are highly degenerated and the parameter space is often non-continuous. We thus prefer to keep the models simple and explore the deviations from a standard, continuous disk rather than to
obtain good fits. Note that in most cases, we expect the structure of the disk to be much more complicated (e.g., containing radial and azimuthal asymmetries, structures within gaps and holes and at the disk edges, etc.) than what is included in our simple models.

Debris disks in the $\eta$\,Cha association are found around late spectral types, from K7 (RECX 1) and M-type stars.
At first approximation, the modeling of the debris disks (see Fig.~\ref{models-dd}) assumes a ring distribution of the debris dust. The SED models allow us to constrain a range of possible locations and ring masses. With masses between 0.1 and 1.8\,$\times$\,10$^{-6}$ M$_\odot$ the ring location varies between 15 and 100 au from the central star. 
The most massive ring, about $\sim 10^{-5}$ M$_\odot$, is found at more than 600 au around RX J1005.3 - 7749. A less massive ring around RECX 1 is also found at very large distances, between 200 and 600 au from the central star. These results are fundamental to further exploring whether the origin of these structures may be related to the presence of an inner solar system. 

\subsection{Dust disk mass versus mass accretion}

\begin{figure}
\centering
\begin{tabular}{c}
\includegraphics[width=0.99\linewidth]{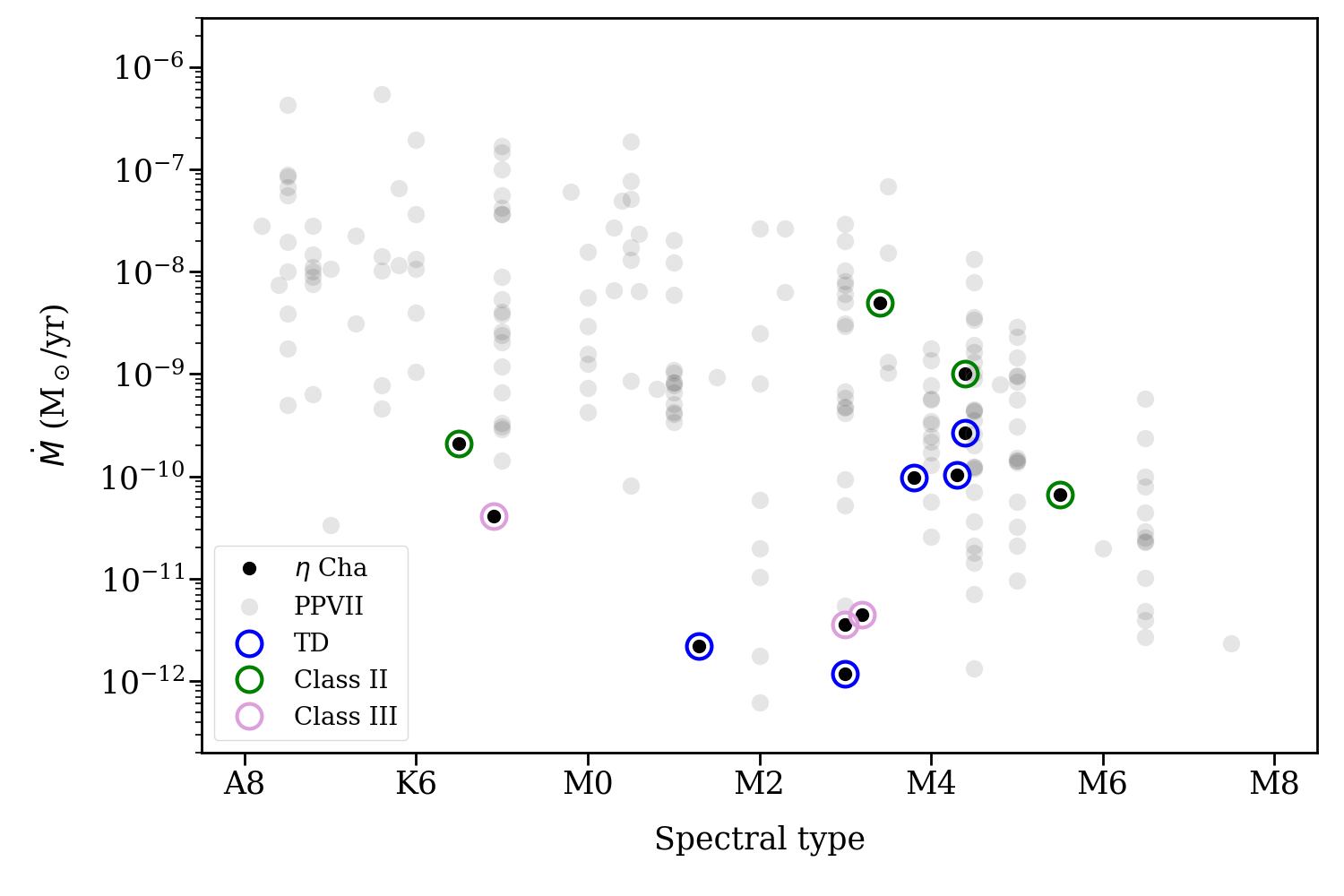}\\
\includegraphics[width=0.99\linewidth]{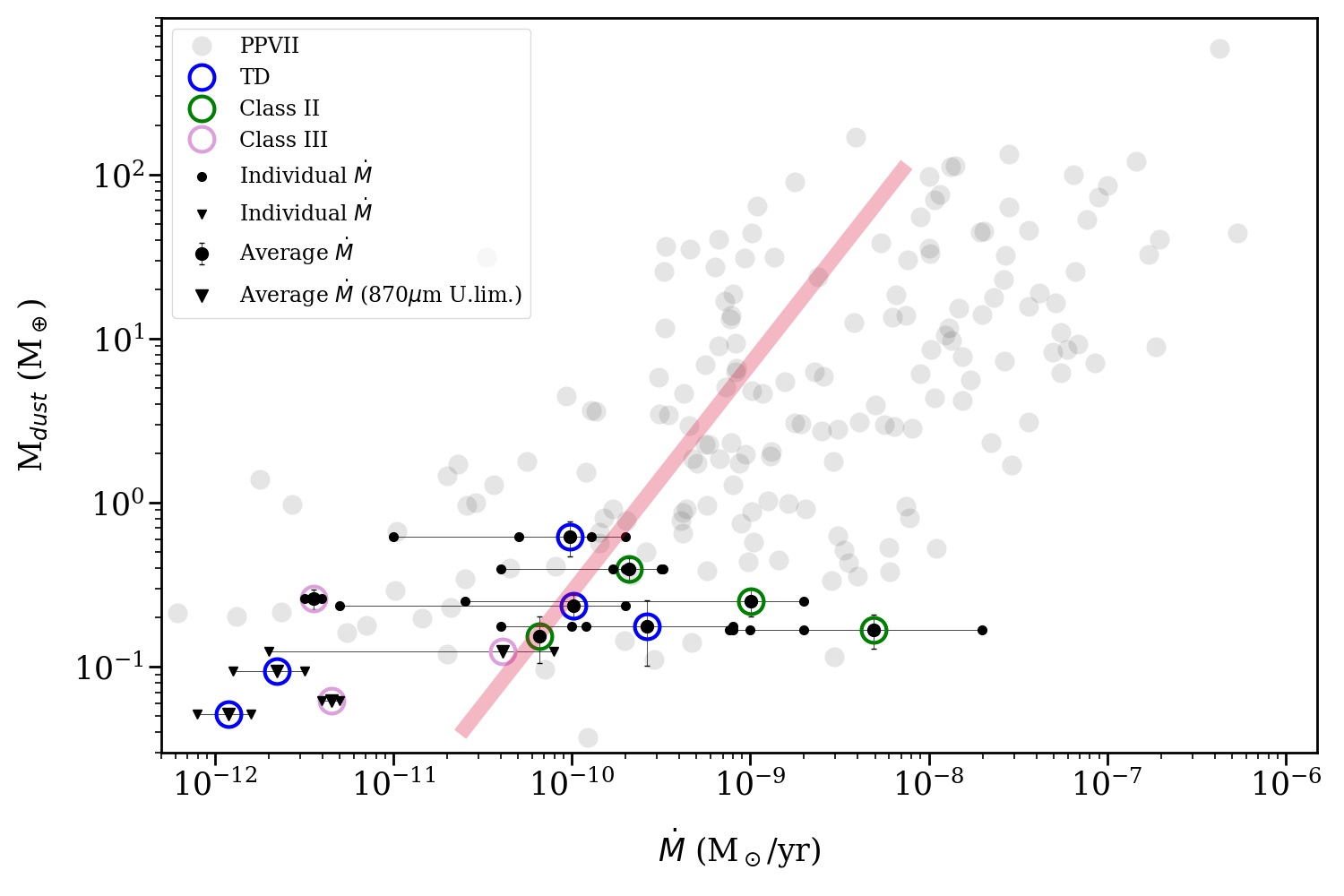} \\
\end{tabular}
\caption{{\it Top}: Accretion rate vs spectral type. Note the lack of significant correlations between these quantities. {\it Bottom}: Circumstellar dust mass derived according to Eqn. \ref{mdust-eq} vs. accretion rate.  For comparison, the objects in the list from \citet{Manaraetal2023} are shown in grey. Note that the total disk mass would depend on the gas-to-dust ratio. APEX flux and mass upper limits are marked by inverted triangles. The red line corresponds to the model fitted by \citet{manara16}. The accretion rates are taken from the literature (see Table \ref{macc}). For each object, we show the individual measurements (small symbols) together with the average rate (large symbols). A further color ring is added to specify the type of disk according to the classical SED classification. \label{mdotmdisk-fig}}
\end{figure}

We first discuss the results on $\eta$\,Cha alone, then in the context of other young sources \citep[compiled by][]{Manaraetal2023} to cover a larger spread in disk dust masses and mass accretion rate.

The relation between submillimeter emission (or disk mass) and accretion has been suggested by several authors, as a means to explore and test viscous accretion and photoevaporation 
\citep[e.g.,][among others]{Hartmann1998,Sicilia-Aguilaretal2013,manara16}. The disks in $\eta$\,Cha have the advantage of being older than most, which allows us to explore whether this relation holds for slow accretors with evolved disks. Table~\ref{macc} lists the mass accretion rates from the literature. To avoid introducing further uncertainties from the various types of disk models that can fit a given SED, and to make our results comparable to those of \citet{manara16}, we derived the
disk mass following Equation \ref{mdust-eq} with the same values they use ($\beta$ = 1, k$_{\nu 0}$ = 10 cm$^2$g$^{-1}$ at $\nu_0$ = 1000 GHz, and T$_{\rm dust}$ = 20 K). Because of the differences in the underlying dust model assumed to obtain the dust absorption coefficient, these masses are different from the masses derived from the RADMC models, which assume a larger maximum grain size and calculate the dust temperature consistently from radiative transfer depending on the disk model and structure. Note that the available data do not constrain the maximum grain sizes, so the differences reflect the typical uncertainties deriving disk masses. 

Spurious correlations may appear because accretion rates and disk masses tend to be correlated with the stellar mass \citep[e.g.,][]{ClarkePringle2006,mulders17}. We thus checked whether this could be the main cause of the marginal correlation, but we found no correlation between the spectral type and the APEX flux (see Fig.~\ref{sptmdisk-fig}) or the accretion rate (see Fig.~\ref{mdotmdisk-fig}). The relation between the accretion rate and the disk mass is shown in the bottom panel of Fig.~\ref{mdotmdisk-fig}. The $\eta$\,Cha data are consistent with the M$_{\rm dust}$ vs \.{M} correlation proposed by \citet{manara16}, although our data points themselves are not significantly or only marginally correlated (when upper limits are included, the false-alarm probability for a Spearman rank test is 7\%). This is likely caused by the large scatter of the mass-accretion relation and the fact that we have very few data points concentrated towards the lower end of mass and accretion rate ranges. 

We also note that there is no evidence in the literature of accretion onto any of the sources that do not have a significant IR excess and that sources with disks classified as Class III \citep[which also correspond to anemic or dust-depleted disks;][]{Ladaetal2006,currie09,sicilia13}, 
have (if any) very low accretion rates. This is the same trend observed in other clusters, where the lack of detectable accretion is strongly correlated with clear signs of disk evolution and substantial mass depletion  \citep[e.g.,][]{sicilia13}.
The presence of very low accretion rates in objects with disks that, albeit evolved and with radial asymmetries, are not cleared from gas and dust in their innermost regions, also poses strong constraints to the photoevaporation rate. The fact that all disks with significant disk mass are accreting is further confirmation that, in order to stop accretion,
a dramatic change in the disk has to occur, and simple holes and gaps are normally not enough if not accompanied by a clear decrease in the disk mass. 

Putting together all observations of accretion, one interesting fact is that accretion towards $\eta$\,Cha members seems to be highly variable, despite its age. Although there are only 2 or 3 measurements per object, 7 out of 12 stars in $\eta$\,Cha have accretion variations of one order of magnitude or more. In contrast, only 10\% of stars in Tr 37 have variable accretion (beyond a factor of few and at least over 3\,$\sigma$) when comparing the existing 2 measurements of a larger collection of members \citep{Sicilia-Aguilaretal2013}. Examining the results in more detail, several of the objects show variable line profiles \citep[e.g.,][]{ingleby13,Lawsonetal2004}. Nevertheless, variability in the line profile does not necessarily indicate variations in the accretion rate and could be instead due to rotational modulation of non-axisymmetric accretion channels \citep{campbellwhite21}. Considering this, together with the fact that detecting very low accretion rates (for which the accretion luminosity is very low compared to the stellar luminosity) is intrinsically very hard, we prefer to be cautious in this respect. 

The $\eta$\,Cha members also suffer the known problem that the observed disk mass (as extrapolated from dust mass, assuming a typical gas-to-dust ratio) is not enough to support accretion for a very long time. In fact, assuming the typical gas-to-dust ratio of 100 and considering the observed accretion rates, all the disks in $\eta$\,Cha would need to entirely disperse within hundreds to thousands of years to support the current accretion rates. Although these relatively old and evolved disks are not expected to live very long, considering the typical evolutionary timescales \citep{Haischetal2001,Sicilia-Aguilaretal2006a,Hernandezetal2007,Fedeleetal2010}, these estimates are too short.
Moreover, all disks are consistent with having their bulk mass at a relatively large distance from the star, which makes it difficult the transport within this short timescale, predicting even shorter lifetimes for disks and accretion. This is statistically implausible because it would mean that all disks would be about to dissipate, in contradiction with the usual smooth disk fraction vs. time curve. Even though the mismatch between accretion and disk masses is a known problem and it tends to affect less the lower-mass end \citep[e.g.,][]{sicilia16}, the data on the $\eta$\,Cha members is a further indication that something is missing in the global picture of accretion and mass transport in disks.

Looking at the young objects from \citet{Manaraetal2023}, Fig.~\ref{mdotmdisk-fig} reveals that our sources are complementary to the sources from the literature. Taking into account the mass accretion variability, mass accretion and disk mass distribution of $\eta$\,Cha does not allow a clear distinction of disk evolutionary path between viscosity, MHD winds, or external photoevaporation as suggested in \citet{Manaraetal2023}.  
We decided on purpose to not specify the age of the sources from  \citet{Manaraetal2023} since those are not directly comparable in any case with an older region as $\eta$\,Cha where there is an intrinsic decreasing of the disk fraction already took place: the surviving accreting sources at such an advanced age would need to be compared only with the same percentage of disks in younger clusters and, as pointed out by \citet{sicilia10}, their younger counterparts would probably correspond to a small subset of those with initially higher disk masses.

\subsection{Disk evolution at longer wavelengths}

The millimeter observations and the age of the cluster allow us to discuss the dissipation of the disks within the first 5 Myr, which represents a key age of the planet formation process since after that age the amount of inventory material is more than halved. 
Our survey includes all the core members of the $\eta$\,Cha associations, not only the Class II sources, plus three halo members from the works of \citet{Murphyetal2010,Murphyetal2011}.
The disk fraction for $\eta$\,Cha in the infrared is based on the {\it Spitzer} work of \citet{Sicilia-Aguilaretal2009} and \citet{fang2013}, including the three halo members.

Since, to a first approximation, the millimeter emission from a protoplanetary disk is tracing its outer region, while the mid-infrared emission traces the inner part, the comparison between the disk fraction in these two wavelength ranges is directly related to the mechanism responsible for the disk dissipation.

We find that all the sources with a near-infrared excess do show a millimeter excess over the stellar photosphere. The only exception is  $\eta$\,Cha itself, which is not a reliable detection as previously discussed (see Section~\ref{etacha-sec}). This suggests that the dissipation processes affect the disk at all radii, as suggested by \citet[][]{Currie&Sicilia-Aguilar}. 

A further discussion of this hypothesis can be done using a recent work of \citet{Michel2021}, where they summarized the updated membership (including recent results from Gaia) and redefined disk classes in a consistent way. Using the definition of disk classes from  \citet{Greene1994} based on the slope in the SED in the 2.2-10~$\mu$m,
 \citet{Michel2021}    derived also the total number of Class I, Class F, and Class II, and the Class III in several young star-forming regions: Corona Australis (hereafter labeled Cr~A), Ophiuchus, Taurus, Chamaeleon I (Cha I), Chamaeleon II (Cha II), IC 348, Lupus, $\epsilon$ Chamaeleontis ($\epsilon$ Cha), TW Hydra (TW Hya), $\eta$\,Cha, and Upper Sco. 
\\
While \citet{Michel2021} analyzed the possible different mechanisms responsible for dust dissipation in disks with and without structures, we compiled instead the fraction of sources classified as Class I, Class F, and Class II and compared them to the fraction of Class III and debris disks. For consistency, we mostly adopt the compiled values from \citet{Michel2021}. However, when comparing disk masses to debris disks \citet{Michel2021} refers to a more complete survey on debris disks carried out by \citet{Hollandetal2017}, where, apart from the TW Hya debris disks, all the others are older than 20~Myrs and most of them isolated field objects. For this reason, we use some additional criteria to compile the final disk fractions which differ from \citet{Michel2021}. The age of $\eta$~Cha is taken from our work, $\epsilon$ Cha from \citet{Murphyetal2010}, TW Hya from \citet{Luhman2023}, and Cr A from \citet{Sicilia-Aguilaretal2011} with a spectroscopic survey of the in-cloud population. The fraction of Class I, F, II, and III for Cr A arises from \citet{peterson11}, which includes only the in-cloud core population of the cluster members spectroscopically characterized. The debris disk fraction for $\eta$\,Cha results from this work, and the one of TW Hya, from the SONS survey \citep[][]{Hollandetal2017}. Following the definition of debris disks used in our work, we also compile the confirmed debris disks from \citet{Michel2021} selecting Class III disks with fractional dust luminosities ${f_{\rm\,dust}\, (\equiv L_{\rm\,dust}/L_{\rm\,*})\,<\, 10^{\,-2}}$.

\begin{figure}
\centering
\begin{tabular}{c}
\includegraphics[width=0.99\linewidth, height = 6.9cm]{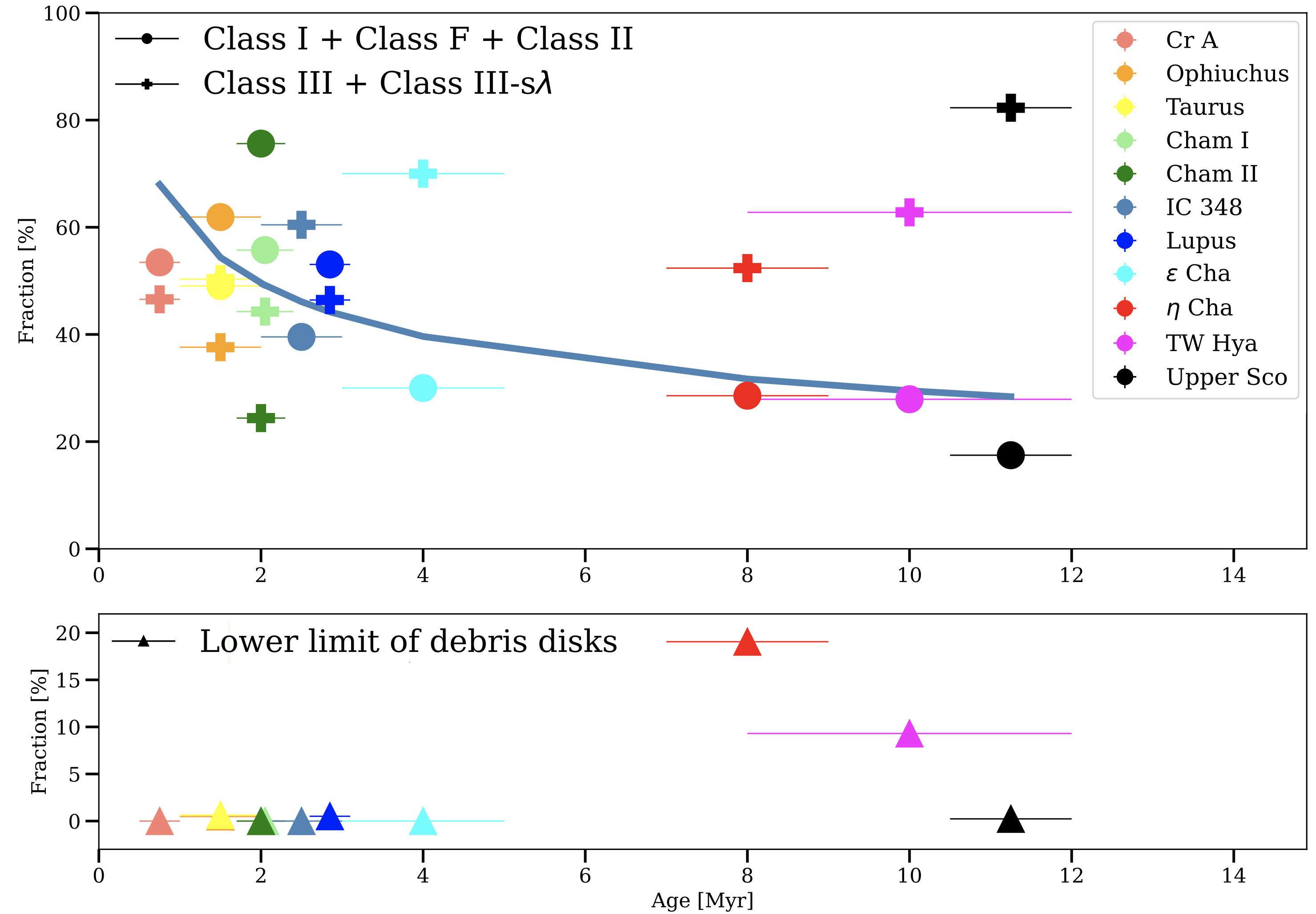} \\
\end{tabular}
\caption{Relation between disk fraction in several young star-forming regions and their age. The fraction of Class I + Class F + Class II disks is marked by filled dots, while the fraction of Class III disks is shown as crosses. Each star-forming region is color-coded according to the legend. The filled triangles in the lower panel represent the lower limits of the debris disk fraction. The blue line represents the fitting of the fraction of Class I + Class F + Class II disks, with a slope of -0.32 $\pm$ 0.11.}
\label{ev-disk}
\end{figure}

Fig.~\ref{ev-disk} shows the fraction of Class I + Class F + Class II disks and Class III disks of young star-forming regions (upper panel), and the lower limit of debris disks in these regions (bottom panel). We find that the fraction of Class III is lower than Class I, F, and II in CrA, Ophiuchus, Cha I, Cha II, and Lupus, while the inverse behavior is seen in Taurus, IC 348, $\epsilon$ Cha, $\eta$\,Cha, TW Hya, and Upper Sco, where the fraction of Class III is higher. We also find an exponential decrease in the fraction of Class I, F, and II with a slope of -0.32 $\pm$ 0.11, with the fraction of such disks decreasing together up to ages 2 to 3 Myrs. After this timescale, the drastic decrease in the disk fractions of Class I, F, and II disks corresponds to an increase in Class III. While this is consistent with previous studies on dust and gas dissipation in low-mass star-forming regions at different ages \citep[e.g.,][]{Hernandezetal2007, Fedeleetal2010,WilliamsCieza2011}, our work also explores whether the percentage of debris disks among Class III objects allow us to further constrain formation and evolution timescales. 

For the younger regions, we find that most of the clusters show only a few percent of debris disks, without allowing proper treatment of the timescale for debris disk formation and evolution. However, our survey on $\eta$\,Cha opens a new discussion on this class of objects showing a percentage of debris disks ten times higher than in other regions. A result in this direction is drawn also including the results of the debris disks in TW Hya from the youngest sources in the SONS survey (explicitly dedicated to debris disks; \citealp{Hollandetal2017}). Furthermore, we find a constant lower limit for debris disks until 8 Myrs, where a much higher percentage of debris disks is found in $\eta$\,Cha. We note that this can be, however, an effect of a bias in the observations of the younger regions that might also host a higher number of debris disk systems, not yet detected.

Nonetheless, this general trend might have some fundamental implications for the timescale of grain growth, dust trapping, and giant planet formation. By 5~Myr, a significant amount of mass will be already as a building block of planets and few sources will have evolved dust up to millimeter sizes. If, instead, a higher fraction of debris disks were to be found in younger regions, it would mean that all these processes can efficiently create stable systems already in a few Myrs.

There are some aspects that need to be taken into consideration, as they may affect the disk fractions. Notably, advancements in observational capabilities, particularly through Gaia \citep[e.g.,][]{Gallietal2018, Roccatagliataetal2018, Herczeg+2019,Roccatagliataetal2020, Luhman2023}, have revealed multiple populations within star-forming regions. To date, the treatment of age and the incorporation of updated disk fractions are only partially addressed. Recent reanalysis of disk fractions with revised membership from Gaia \citep[e.g.,][]{Mendigutia+2022} recover consistent disk fractions (with discrepancies up to 10\%),  despite considering a more extensive area of star-forming regions. Combining different techniques to complete the cluster membership in different spectral type ranges, as done by \citet{Pelayo-Baldarragoetal2023}, also found that the disk fraction in IC 1396 decreased from 39\% down to 28$\%$ at about 4 Mys.

\section{Conclusions and Summary}
\label{conclusions}

The main results and conclusions of our study can be summarized as follows.
\begin{itemize}
\item 
Thanks to its proximity, our pilot submillimeter survey carried out with the APEX/LABOCA instrument detected 12 (two of them marginally) out of the 20 members of the $\eta$\,Cha cluster, providing valuable upper limits for the rest. Our work also confirms the distance of the $\eta$\,Cha association to be at  99.0 $\pm$ 1.3 pc, with an age between 7 and 9 Myr. 
\item 
The modeling of the SEDs, including the new submillimeter data, suggests that protoplanetary disks in $\eta$\,Cha have holes with radii between 0.01 and 0.03 AU, while ring-like emission from the debris disks arises from between 20 au and 650 au from the central star. In the case of ECHA J0801, both protoplanetary disk and debris disk models led to a large gap of about 15 au and a very massive disk compared with other systems.
We confirm the trend that low disk masses and low accretion rates are correlated, as well as the fact that having gaps and holes is not enough to stop disk accretion, if not accompanied by a substantial decrease in the disk mass. 
\item 
All disks in the cluster depart from the typical structure of primordial protoplanetary disks, showing evidence of radial changes in their structures. Although most disks observed at high resolution have signatures of radial and azimuthal asymmetries (so that their presence may not be representative of a rapid evolutionary stage, but may also correspond to long-lived features), for $\eta$\,Cha disks these asymmetries are already evident in the SED, which indicates a further degree of evolution. Disk dispersal is found to be dominated by inside-out processes.
\item 
Comparing accretion rates and disk masses, the classical problem of lack of disk matter to support accretion is also present in $\eta$\,Cha, extending well into the low-mass regime where the problem tends to be less serious than for intermediate-mass stars. 
\item 
Our work on the $\eta$ Cha association helps put in context the evolution of disks by addressing all the different types of disks and their changes with age. An important point is that we include the debris disks, for which so far there is only a lower limit. Our study shows consistency with previous results concerning the fractions of Class I, F, Class II, and Class III objects, while also bringing up the discussion on the formation and evolution of debris disks. Their higher fraction in older clusters suggests that there are physical mechanisms capable of efficiently forming these systems in less than 5 Myr. 
\end{itemize}

\begin{acknowledgements}

VR acknowledges the support of the Italian National Institute of Astrophysics (INAF) through the INAF GTO Grant  “ERIS $\&$ SHARK GTO data exploitation”. 
ASA and JCW were in part supported by the STFC grant number ST/S000399/1 (``The Planet-Disk Connection: Accretion, Disk Structure, and Planet Formation"). JCW is funded by the European Union under the European Union’s Horizon Europe Research $\&$ Innovation Programme 101039452 (WANDA). MK and SW acknowledge the DFG for financial support under contract WO 857/15-1 and WO 857/15-2 within the context of the Research UnitFOR 2285 ``Debris Disks in Planetary Systems''. MK also acknowledges the funding from the Royal Society. This publication is based on data acquired with the Atacama Pathfinder Experiment (APEX). APEX is a collaboration between the Max-Planck-Institut f\"ur Radioastronomie, the European Southern Observatory, and the Onsala Space Observatory. This research has made use of the SIMBAD database, operated at CDS, Strasbourg, France.
\end{acknowledgements}
\begin{appendix}
 
\section{Additional Material}
In this Appendix, we present the extended Tables of all the data used during the discussion in the main paper. 
\begin{table*}
\begin{small}
\caption{Mass accretion rates compiled from the literature. The accretion rates labeled as \.{M}$_{\rm model}$ are derived from magnetospheric accretion model 
fits, while rates marked as \.{M}$_{\rm V10}$ are obtained from the relation between accretion rate and the H$\alpha$ line width at 10\% of the maximum (V10), from \citet{Nattaetal2004}. The uncertainties in the disk mass include detection and calibration uncertainties. ECHA 0841 has TO/flat SED, EW(Ha) = -12 AA. ECHA 0844 has flat SED and is likely accreting with EW(Ha) = -58 AA 
{\it References:} J06: \citet{Jayawardhanaetal2006}; 
L04: \citet{Lawsonetal2004}; 
M11: \citet{Murphyetal2013}; 
I13: \citet{ingleby13};
R18: \citet{Rugeletal2018}; 
* = V10 below 200 km s$^{-1}$, which is the minimum for the \citet{Nattaetal2004} relation.} 
\label{macc}      
\centering               
\begin{tabular}{l r r r r }  
\hline\hline                
\noalign{\smallskip}
Source		&	log(\.{M}$_{\rm model}$)		& Refs. &	log(\.{M}$_{\rm V10}$)	&	Refs.  \\
		&   (M$_\odot$/yr) 			&   	& 	(M$_\odot$/yr) 	\\
\noalign{\smallskip}
\hline                
\noalign{\smallskip}
RECX-1 			&	    -11.9		&	J06	&				&				            \\
RECX-3			&	    -12.1*		&	L04 	&	-11.8*			&	J06		\\
RECX-4			&	    -11.9*		&	L04	&	-11.5*			&	J06	\\
RECX-5			&	    -10.3		&	L04	& 	-9.7,-11.0*, -9.89	&    L04, J06, R18	\\
RECX-6			&	    -11.4*		&	L04 	&	-11.5*			&	J06	\\
RECX-7			&	    -11.7*		&	L04	&	-10.1 			&	J06	\\
RECX-9			&	-10.4			&	L04	& -10.0,-9.1, -9.92	 	&   L04, J06, R18\\
RECX-10			&				&	   	&  -11.8*, -11.9* 		&	L04, J06\\
RECX-11			&	-10.4,-9.77		&	L04,I13	& -9.7,	-9.5, -9.49 	&  L04, J06, R18  \\
RECX-12			&				&	   	&  -11.3*, -11.4*		&	L04,J06	\\
ECHA\,J0843 		&	-9.0,-9.10 		& L04, I13 & -7.7,-8.7,-9.12	&  L04 , J06\\
ECHA\,J0844 		&	 			&		&       -10.18  		&      	R18		\\
2MASS\,J0801		&	-11.3*, -9.7		&   M11		&			&				 \\
2MASS\,J0820		&	-10.6, -9.7		&   M11		&	-8.7			&	M11		    \\
 \noalign{\smallskip}
\hline                        
\noalign{\smallskip}
\end{tabular}
\end{small}
\end{table*}

\begin{table}
\begin{small}
\caption{Log of the observations of the $\eta$ Cha members observed in on-off mode.}
\label{log_1}      
\centering               
\begin{tabular}{l c c c}  
\hline\hline                
\noalign{\smallskip}
Source      &  Obs. Date & Scan \#& Exp. time [s]\\
\noalign{\smallskip}
\hline                
\noalign{\smallskip}
$\eta$\,Cha      &2010-10-30&81846&600\\
		 &&81847&600\\
		 &&81854&600\\
\noalign{\smallskip}
\hline                
\noalign{\smallskip}
J0801        &2010-12-11&91881&600\\
		 &&91882&600\\
		 & &91883&600\\
		 & &91884&600\\
		 & &91885&600\\
		 & &91886&600\\
		 & &91887&600\\
\noalign{\smallskip}
\hline                
\noalign{\smallskip}
J0841        &2010-12-10&91567&600\\
		&&91568&\\
	         &&91569&\\
	         &&91570&\\
	         &&91571&\\
	         &&91572&\\
\noalign{\smallskip}
\hline                
\noalign{\smallskip}
J0843	&2010-12-10&81872&600\\
	         &&81873&\\
	         &&96114&\\
	         &&96115&\\
	         &&96116&\\
	         &&96117&\\
	         &&96118&\\
\noalign{\smallskip}
\hline                
\noalign{\smallskip}
J0844       &2010-12-10&91697&600\\
	        &&91698&\\
        		&&91699&\\
        		&2010-12-10&91700&600\\
	        &2010-12-27&96127&600\\
	        &&96128&\\
	        &&96129&\\
	        &&96130&\\
	        &&96131&\\
\noalign{\smallskip}
\hline                
\noalign{\smallskip}
RECX-3    &2010-12-26&95914&600\\
		&&95915&\\
		&&95916&\\
		&&95917&\\
		&&95918&\\
		&&95919&\\
		&&95920&\\
\noalign{\smallskip}
\hline                
\noalign{\smallskip}
RECX-4    &2010-12-26&95928&600\\
		&&95929&\\
		&&95930&\\
		&&95931&\\
		&&95932&\\
		&&95933&\\
\noalign{\smallskip}
\hline                
\noalign{\smallskip}
RECX-5 	&	2010-10-30 	&	81865	&	600	\\
	&		&	81867	&		\\
	&		&	81868	&		\\
\noalign{\smallskip}
\hline                
\noalign{\smallskip}

\end{tabular}
\end{small}
\end{table}

\begin{table}
\begin{small}
\caption{Journal of the APEX/LABOCA observations of $\eta$ Cha members obtained in on-off mode.}
\label{log_2}      
\centering               
\begin{tabular}{l c c c}  
\hline\hline                
\noalign{\smallskip}
Source      &  Obs. Date & Scan \#& Exp. time [s]\\
\noalign{\smallskip}
\hline                
\noalign{\smallskip}
RECX-6 &   2013-11-18   &   82368 & 360   \\
&      &   82369 & 600   \\
  &      &   82370 & 600   \\
  &      &   82371 & 600   \\
  &      &   82372 & 600   \\
  &      &   82373 & 60    \\
  &   2013-11-21   &   83354 & 600   \\
  &      &   83355 & 600   \\
  &      &   83356 & 600   \\
  &      &   83357 & 600   \\
\noalign{\smallskip}
\hline                
\noalign{\smallskip}
RECX-7 &   2013-11-19   &   82822 & 600  \\
 &      &   82823 & 600   \\
  &   2013-11-20   &   83105 & 600   \\
  &     &   83106 & 600   \\
\noalign{\smallskip}
\hline                
\noalign{\smallskip}
RECX-9    &2010-10-30&81869&600\\
		&&81870&\\
		&&81871&\\
 &   2013-11-19  &   82824 & 600   \\
 &      &   82825 & 600   \\
  &   2013-11-20   &   83103 & 600  \\
  &      &   83104 (no) & 600   \\
\noalign{\smallskip}
\hline                
\noalign{\smallskip}
RECX-11  &2010-10-30&81858&600\\
		&&81859&\\
		&&81860&\\
 \noalign{\smallskip}
\hline                
\noalign{\smallskip}
RECX-12 &   2013-11-18   &   82347 & 600  \\
 &      &   82348 & 600   \\
 &      &   82349 & 570   \\
 &      &   82350 & 600   \\
 &      &   82351 & 600   \\
 &      &   82352 & 330   \\
 &   2013-11-21   &   83361 & 600   \\
 &      &   83362 & 600   \\
\noalign{\smallskip}
\hline                        
\noalign{\smallskip}
RS Cha &   2013-11-18   &   82375 & 600   \\
  &      &   82376 & 600   \\
\noalign{\smallskip}
\hline                        
\noalign{\smallskip}
HD75505&   2013-11-19   &   82820 & 600   \\
&      &   82821 & 600  \\
\noalign{\smallskip}
\hline                        
\noalign{\smallskip}
\end{tabular}
\end{small}
\end{table}

\begin{table}
\begin{small}
\caption{Journal of the observations of the $\eta$ Cha members in mapping mode (we report only the scans that have been analyzed with more than one sub-scan.)}
\label{log_map}      
\centering               
\begin{tabular}{l c c c}  
\hline\hline                
\noalign{\smallskip}
Source      &  Obs. Date & Scan \#& Exp. time [s]\\
\noalign{\smallskip}
\hline                
\noalign{\smallskip}
RECX-16 & 2008-10-21 & 55435 & 35\\ 
&  & 55436 & 35 \\
&  & 55437 & 35 \\
&  & 55438 & 35 \\
&  & 55439 & 35 \\
&  & 55440 & 35 \\
&  & 55441 & 35 \\
&  & 55442 & 35 \\
\noalign{\smallskip}
\hline                
\noalign{\smallskip}
RECX-17  & 2008-12-26 & 73721 & 420 \\
 &  & 73722 & 420 \\
 &  & 73723 & 420 \\
 &   & 73724 & 420 \\
 &   & 73725 & 420 \\
 &   & 73726 & 420 \\
 &   & 73727 & 420 \\
 &   & 73728 & 420 \\
\noalign{\smallskip}
\hline                
\noalign{\smallskip}
RECX-18 & 2008-12-26 & 73697 & 420 \\
 &   & 73698 & 420 \\
 &   & 73699 & 420 \\
 &   & 73700 & 420 \\
 &   & 73701 & 420 \\
 &   & 73702 & 420 \\
 &   & 73703 & 420 \\
 &   & 73704 & 420 \\
 \noalign{\smallskip}
\hline                        
\noalign{\smallskip}
\end{tabular}
\end{small}
\end{table}

\clearpage
\onecolumn
\begin{landscape}
\begin{table}
\begin{small}
\caption{Multi-wavelength photometry including the near- and mid-infrared data from \citet{Sicilia-Aguilaretal2009} and far-infrared from \citet{Riviere-Marichalaretal2015}.}
\label{photometry}
\begin{longtable}{l r|r|r|r|r|r|r|r|r|r|r}  
\hline\hline                
\noalign{\smallskip}
 \noalign{\smallskip}
Name	&	2MASS ID	&	3.6 $\mu$m	&	4.5 $\mu$m	&	5.8 $\mu$m	&	8.0 $\mu$m	&	24 $\mu$m	&	70 $\mu$m	&	70 $\mu$m	&	100 $\mu$m	&	160 $\mu$m	&	870 $\mu$m	\\
\noalign{\smallskip}
 \noalign{\smallskip}
 &		&							{\it Spitzer}/IRAC	&	IRAC	&	IRAC	&	IRAC	&	{\it Spitzer}/MIPS	&	MIPS	&	{\it Herschel}/PACS	&	PACS	&	PACS	&	LABOCA	\\
\hline\hline                
\noalign{\smallskip}
 \noalign{\smallskip}								
$\eta$\,Cha	&	08411947-7857481	&	$^a$	&	$^a$	&	$^a$	&	$^a$	&	4.33 $\pm$ 0.02	&	3.87 $\pm$ 0.13	&	$<$6	&		&	$<$16	&	3.37$\pm$1.86	\\
RS Cha	&	08431222-7904123	&	$^a$	&	$^a$	&	5.40 $\pm$ 0.01	&	5.40 $\pm$ 0.01	&	5.34 $\pm$ 0.03	&		&		&		&		&	18.62$\pm$2.63	\\
HD 75505	&	08414471-7902531	&	6.97 $\pm$ 0.01	&	6.97  $\pm$ 0.01	&	6.97 $\pm$ 0.01	&	6.97 $\pm$ 0.01	&	6.97 $\pm$ 0.06	&		&	$<$7	&		&	$<$19	&	$$<$$7.86	\\
J0836	&	08361072-7908183	&	10.57  $\pm$ 0.03 	&	10.45  $\pm$ 0.01 	&	10.40 $\pm$ 0.01 	&	10.37 $\pm$ 0.01 	&	$^b$	&	$^b$	&	$<$10	&		&	$<$24	&		\\
J0838	&	08385150-7916136	&	10.07  $\pm$ 0.02 	&	10.00  $\pm$ 0.01 	&	9.92 $\pm$ 0.01	&	9.90 $\pm$ 0.01	&	$^b$	&	$^b$	&	$<$8	&		&	$<$15	&		\\
J0841	&	08413030-7853064	&	10.53  $\pm$ 0.01 	&	10.26  $\pm$ 0.05 	&	10.04 $\pm$ 0.01 	&	9.48 $\pm$ 0.01	&	7.07 $\pm$ 0.06	&		&	$<$15	&	$<$12	&	$<$25	&	2.88$\pm$0.97	\\
J0843	&	08431857-7905181	&	8.38  $\pm$0.03	&	7.91  $\pm$ 0.01	&	7.42 $\pm$ 0.03	&	6.51 $\pm$ 0.01	&	3.52 $\pm$ 0.01	&	1.89 $\pm$ 0.02	&	204$\pm$6 & 142$\pm$5	&	71$\pm$7	&	5.42$\pm$0.83	\\
J0844	&	08440914-7833457	&	11.02 $\pm$0.01 	&	10.75  $\pm$ 0.01 	&	10.42 $\pm$ 0.01 	&	9.76 $\pm$ 0.01	&	7.11 $\pm$ 0.07	&	$^b$	&	29$\pm$4	&		&	59$\pm$8	&	4.96$\pm$0.83	\\
RECX-1	&	08365623-7856454	&	7.17  $\pm$ 0.02	&	7.20  $\pm$ 0.02	&	7.16 $\pm$ 0.01	&	7.11 $\pm$ 0.02	&	6.96 $\pm$ 0.06	&		&		&		&		&		\\
RECX-3	&	08413703-7903304	&	9.27  $\pm$ 0.02	&	9.21  $\pm$ 0.02	&	9.09 $\pm$ 0.05	&	9.15 $\pm$ 0.01	&	8.49 $\pm$ 0.13	&		&	6$\pm$1$^{\star\star}$	&	$<$10	&	$<$13	&	$<$1.65	\\
RECX-4	&	08422372-7904030	&	8.45  $\pm$ 0.02	&	8.45  $\pm$ 0.02	&	8.37 $\pm$ 0.01	&	8.32 $\pm$ 0.02	&	7.73 $\pm$ 0.09	&		&	9$\pm$1$^{\star\star}$ 	&	$<$9	&	$<$14	&	$<$3.03	\\
RECX-5	&	08422710-7857479	&	9.59  $\pm$ 0.03	&	9.50  $\pm$ 0.03	&	9.37 $\pm$ 0.01	&	8.89 $\pm$ 0.01	&	5.05 $\pm$ 0.02	&	2.38 $\pm$ 0.04	&	113$\pm$24	&	203$\pm$13	&	138$\pm$28	&	20.01$\pm$2.45	\\
RECX-6	&	08423879-7854427	&	9.15$\pm$ 0.04	&	9.07  $\pm$ 0.07	&	9.04$\pm$ 0.01	&	9.04 $\pm$ 0.01	&	8.88 $\pm$ 0.16	&		&	$<$5	&		&	$<$13	&	8.38$\pm$0.54	\\
RECX-7	&	08430723-7904524	&	7.52$\pm$ 0.01	&	7.54  $\pm$ 0.02	&	7.49$\pm$ 0.01	&	7.46 $\pm$ 0.01	&	7.37 $\pm$ 0.08	&		&	$<$15	&	$<$11	&	$<$21	&	$<$4.02	\\
RECX-9	&	08441637-7859080	&	8.99$\pm$ 0.02	&	8.80  $\pm$ 0.01	&	8.57$\pm$ 0.01	&	7.97 $\pm$ 0.02	&	5.8 $\pm$ 0.03	&	3.16$\pm$0.07	& &	77$\pm$3	&	53$\pm$7	&	$<$5.46	\\
RECX-10	&	08443188-7846311	&	8.58$\pm$ 0.01	&	8.61  $\pm$ 0.01	&	8.55$\pm$ 0.02	&	8.49 $\pm$ 0.01	&	8.45 $\pm$ 0.13	&		&	$<$7	&		&	$<$14	&		\\
RECX-11	&	08470165-7859345	&	7.09$\pm$ 0.02	&	6.86  $\pm$ 0.01	&	6.57$\pm$0.01	&	5.97$\pm$0.01	&	3.68$\pm$0.01	&	1.80$\pm$0.02	&	212$\pm$21	&	205$\pm$13	&	174$\pm$26	&		\\
RECX-12	&	08475676-7854532	&	8.19$\pm$0.01	&	8.15 $\pm$ 0.01	&	8.10$\pm$ 0.02	&	8.07$\pm$0.01	&	7.95$\pm$0.10	&		&	$<$5	&		&	$<$13	&	$<$2.01	\\ 
\hline                        
\noalign{\smallskip}
\footnote{$^{\star\star}$:\citet{ciezaetal2013}.}
\end{longtable}
\end{small}
\end{table}
\end{landscape}

\twocolumn

\begin{figure}[h!]
\centering
\includegraphics[width=9cm]{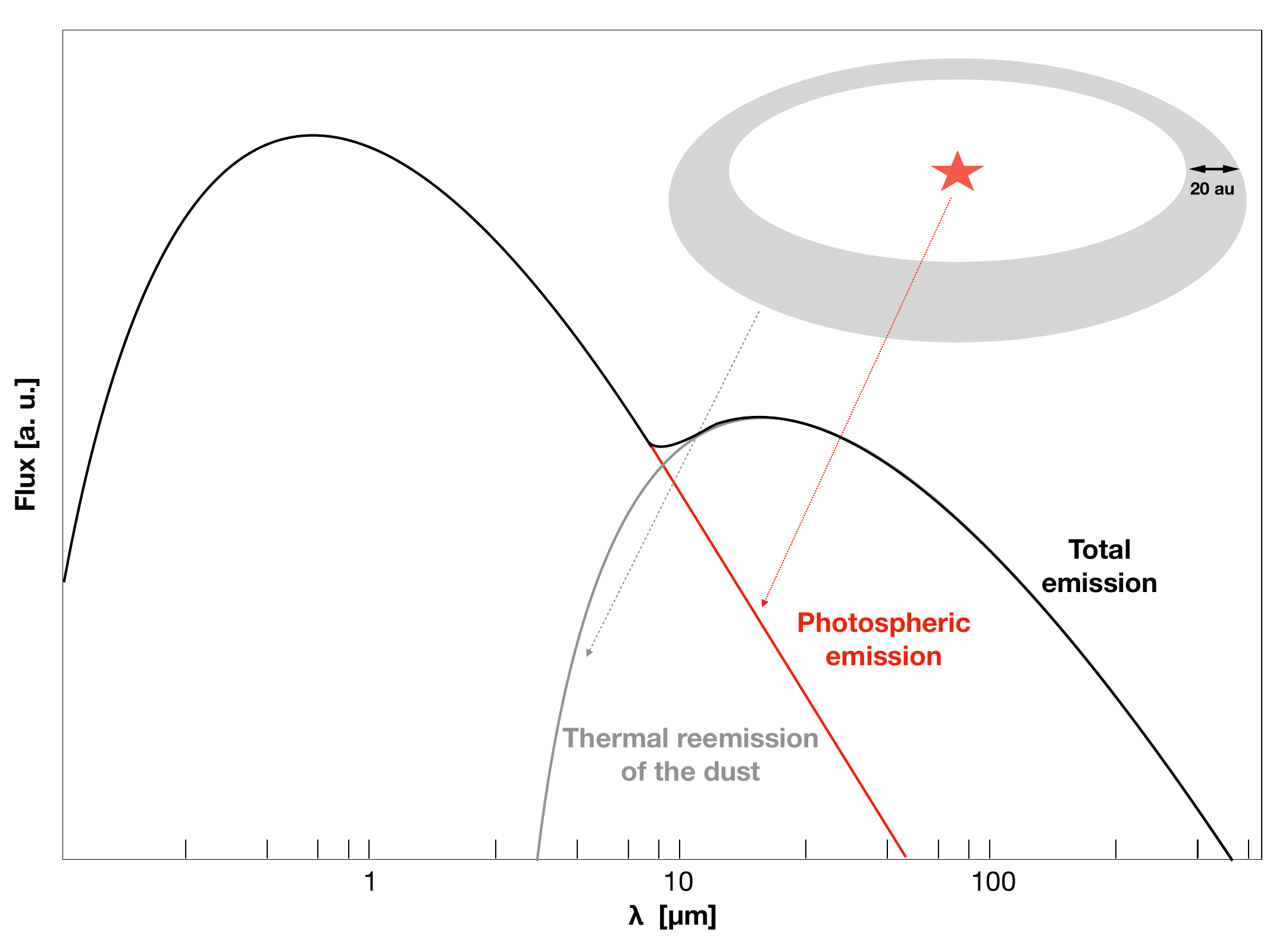}
\caption{Sketch of the basic idea of the model used to interpret the SED of the debris disks.} 
\label{figVibStab}
\end{figure}

We show here the results of the tentative interpretation of the SEDs of RECX 6 and RECX 8. Both sources are well detected with LABOCA at 870 $\mu$m above 3\,$\sigma$. 

\begin{figure*}
\centering
\begin{tabular}{cc}
\includegraphics[width=0.4\linewidth]{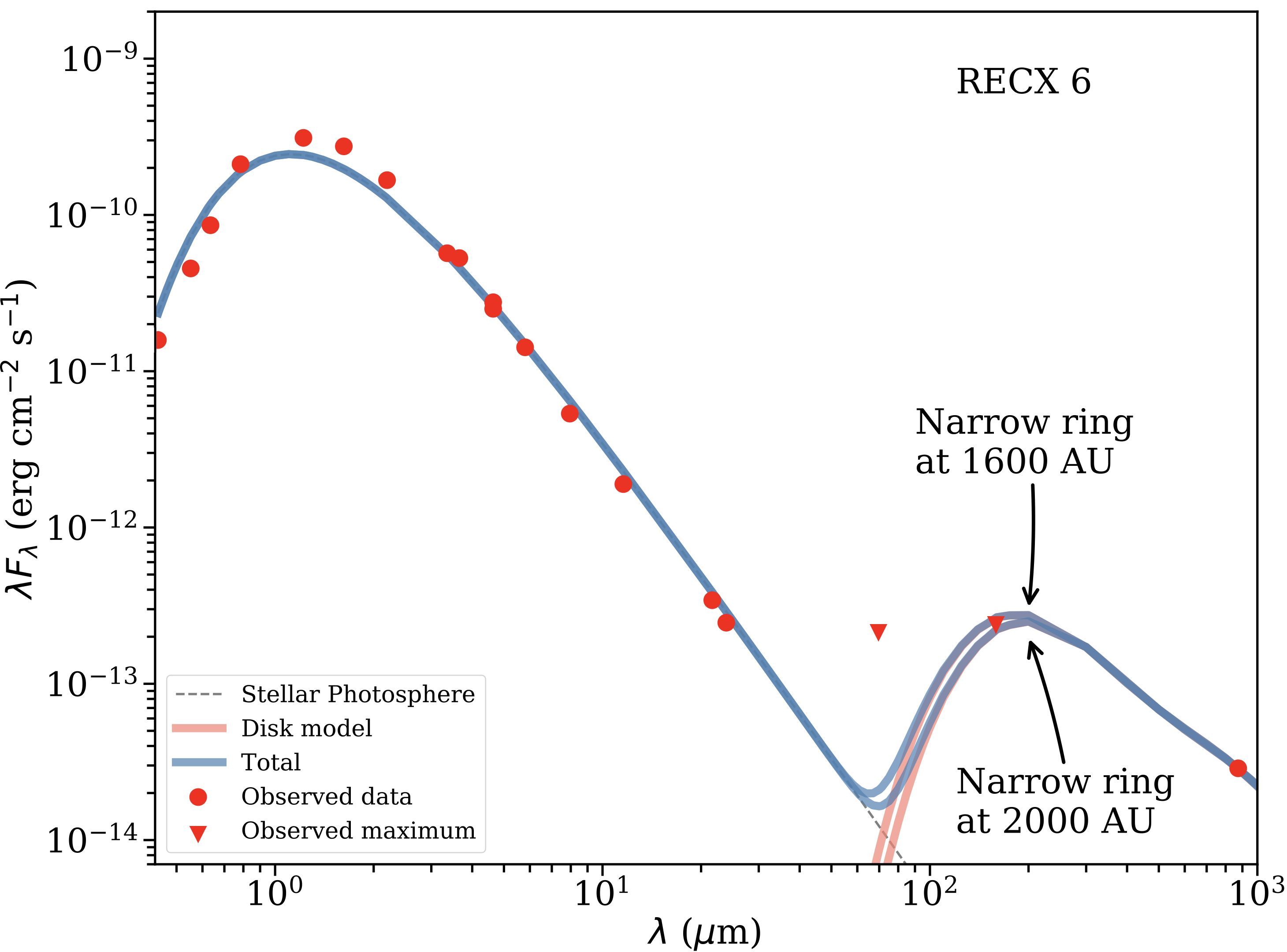} &
\includegraphics[width=0.4\linewidth]{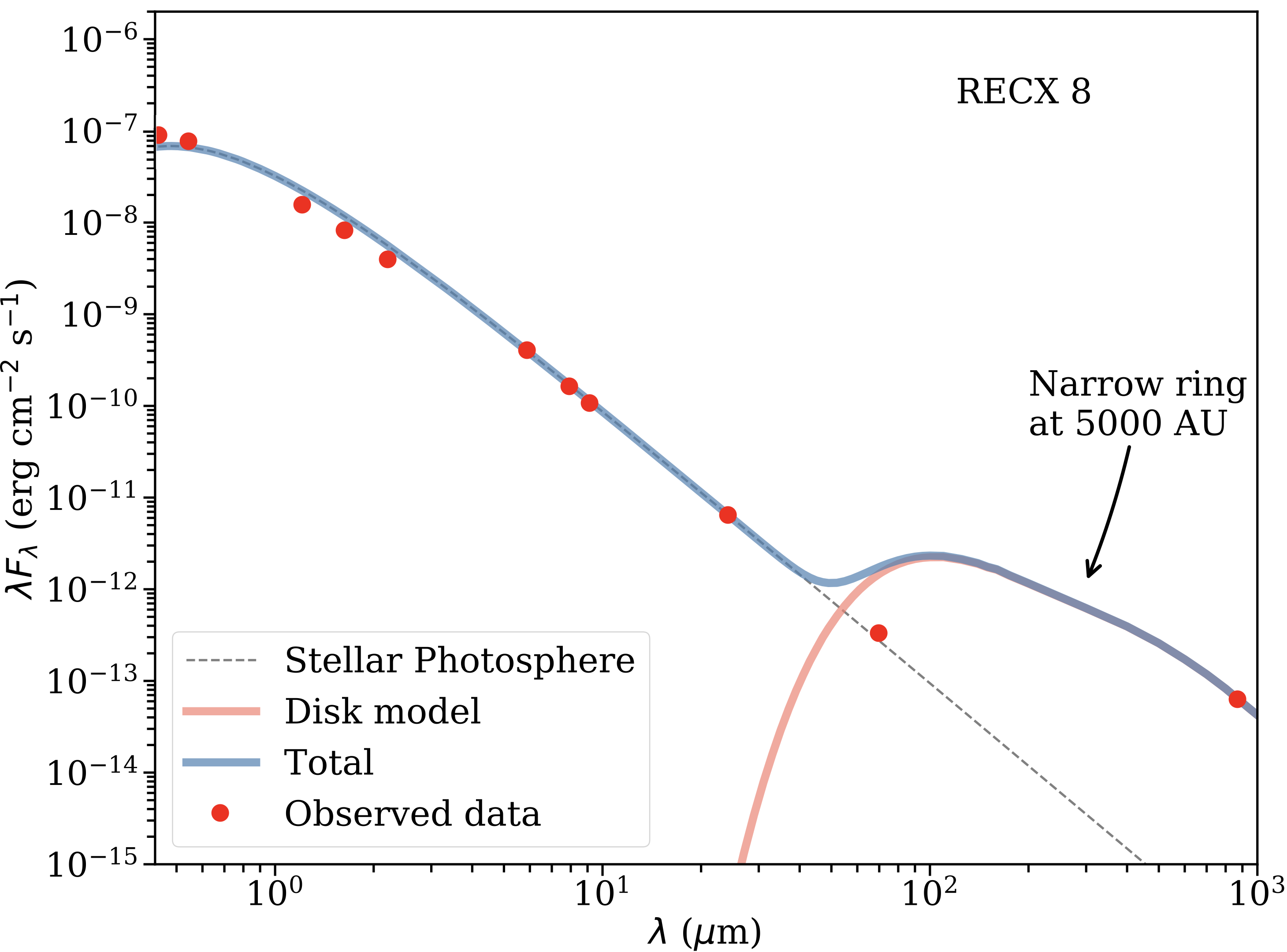} \\
\end{tabular}
\caption{Tentative SED models for the debris disks. The photometry data are marked by red circles and red triangles (the upper limits).  Stellar photospheric, disk, and total emission (stellar photospheric + disk models) are represented by grey dotted lines, and red and blue continuous lines, respectively, but none of those are able to reproduce all the observed data well. \label{models-ddbad}}
\end{figure*}

\end{appendix}

\bibliographystyle{aa}
\bibliography{references}
\end{document}